\documentclass[article,notitlepage,preprintnumbers,superscriptaddress,nofootinbib,secnumberarabic,tightenlines,11pt]{revtex4-2}
\usepackage[utf8]{inputenc}
\usepackage[a4paper, total={7in, 9in}]{geometry}
\usepackage{amsmath}
\usepackage{amssymb}
\usepackage{enumerate}
\usepackage{amsmath}
\usepackage{tikz}
\usepackage[compat=1.1.0]{tikz-feynman}
\usepackage{feynmf}
\usepackage{slashed}
\usepackage{braket}
\usepackage{url}
\usepackage{natbib}
\usepackage{graphicx}
\usepackage[pdftex,bookmarks,linktocpage,pdfpagelabels,plainpages=false,hyperfigures,linkcolor=cyan,citecolor=cyan,urlcolor=magenta]{hyperref} 
\hypersetup{colorlinks=true}
\usepackage[capitalize]{cleveref}
\usepackage{mathrsfs,amssymb}  
\usepackage{cancel}
\usepackage[normalem]{ulem}

\usepackage[thinc]{esdiff}
\usepackage{fontawesome}
\usepackage{xcolor}
\usepackage{xspace}
\usepackage{appendix}

\newcommand{\mpbh}{\ensuremath{M_{\rm PBH}}}
\newcommand{\fgw}{\ensuremath{f_{\rm GW}}}
\newcommand{\fpbh}{\ensuremath{f_{\rm PBH}}}
\newcommand{\hto}{\ensuremath{h^2 \Omega_{\rm GW}}}

\newcommand{\rsep}{\ensuremath{R_{\rm sep}}}

\makeatletter
\def\@hangfrom@section#1#2#3{\@hangfrom{#1#2}#3}
\def\@hangfroms@section#1#2{#1#2}
\makeatother

\makeatletter
\def\p@subsection{}
\def\p@subsubsection{}
\makeatother

\usepackage{titlesec}
\titleformat{\section}[hang]{\Large\bfseries}{\thesection}{1em}{}
\titleformat{\subsection}[hang]{\large\bfseries}{\thesubsection}{1em}{}
\renewcommand{\thesection}{\arabic{section}}
\renewcommand{\thesubsection}{\thesection.\arabic{subsection}}

\newcommand{\vev}{$\langle\Phi\rangle$\xspace}
\newcommand{\vevMth}{\langle\Phi\rangle}

\begin{document}
\preprint{MI-HET-886}
\preprint{CETUP2026-01}
\author{Bhaskar Dutta}
\email{dutta@tamu.edu}
\affiliation{Texas~A$\&$M~University,~College~Station,~TX~77843,~USA}

\author{Cash Hauptmann}
\email{chauptmann2@huskers.unl.edu}
\affiliation{Department of Physics and Astronomy, University of Nebraska, Lincoln, NE 68588}

\author{Peisi Huang}
\email{peisi.huang@unl.edu}
\affiliation{Department of Physics and Astronomy, University of Nebraska, Lincoln, NE 68588}

\author{Adrian Thompson}
\email{a.thompson@northwestern.edu}
\affiliation{Northwestern~University Department of Physics and Astronomy,~Evanston,~IL~60208,~USA}

\title{PBH formation and Gravitational Waves as Multi-messenger Signals \\of First-order Phase Transitions}

\begin{abstract}
    The collapse of false-vacuum domains during first-order phase transitions in the early Universe may lead to primordial black hole (PBH) formation whose signatures form a multimessenger complement to gravitational wave (GW) production.
    We focus on PBH formation through the gravitational collapse of false-vacuum domains, described using a junction condition formalism.
    This formalism develops the Schwarzschild collapse criterion dynamically, avoiding the usage of critical overdensity thresholds in a post-inflationary Universe, and is driven solely by the vacuum energy enclosed within shrinking false-vacuum domains without the assistance of particle or domain wall interactions in the false vacuum.
    We study the parameter space of phase transitions and identify regions producing observable GWs, observable PBHs, or both simultaneously. We investigate this phenomenology in polynomial and classically conformal scalar field potentials as model benchmarks.
    We find that the scalar fields with vacuum expectation values $\vevMth$ in the range of 1-100 MeV have the largest model parameter space available for these multi-messenger signals,
    which are testable with upcoming GW observatories, searches for Hawking radiation, and gravitational lensing surveys.
\end{abstract}

\maketitle
\tableofcontents  

\section{Introduction}

Many extensions of the Standard Model (SM) predict cosmological first-order phase transitions (FOPTs), which provide distinct astrophysical probes of particle physics at energies inaccessible to terrestrial experiments. The nucleation, expansion, and subsequent collision of true-vacuum domains during these transitions can generate a stochastic background of gravitational waves (GWs) that may be detected by current and future GW observatories.
FOPTs may also produce primordial black holes (PBHs), opening complementary channels of observations.
PBHs with smaller masses may emit observable fluxes of Hawking radiation~\cite{Page:1976wx, Carr:1998fw}, while heavier ones could be detected via gravitational lensing~\cite{Carr:2020gox}.
If the PBHs have not completely evaporated in the present epoch, they are expected to make up some fraction of the dark matter and are then subject to abundance constraints~\cite{Carr:2020gox}.
These signatures together offer a multi-messenger probe of the underlying particle physics responsible for the FOPT and motivate a detailed understanding of how the microscopic properties of a model determine its correlated GW and PBH phenomenology.

While the production of GWs in FOPTs is well-understood, several distinct mechanisms of forming PBHs during these transitions have been proposed.
These include collapse triggered by overdensities from delayed vacuum transitions, which become significant mainly for slow and moderately strong FOPTs~\cite{Kawana:2022olo};
collapse of the latest-percolating patches in sufficiently supercooled transitions, typically requiring FOPTs lasting more than 15\% of a Hubble time~\cite{Gouttenoire:2023naa};
and PBH production from collisions of unusually early bubbles with macroscopically thick walls~\cite{Jung:2021mku}.
Many other mechanisms involve the trapping of matter within contracting pockets of the old phase (false vacuum): if the overdensity of matter within these false vacuum domains exceeds a compactness condition, PBHs can form directly~\cite{Gross:2021qgx, Baker:2021nyl, Baker:2021sno} or through intermediate Fermi-ball, or Q-ball states~\cite{Kawana:2021tde, Hong:2020est,Huang:2022him,Dent:2025lwe}. These mechanisms typically require either a pre-existing asymmetry in the (dark) fermion sector that promotes collapse, or a narrow range of Yukawa couplings between the fermions and the scalar field in order to prevent back-pressure during compression and maintain heavy fermion masses in true vacuum to trap them within the false-vacuum patch~\cite{Hong:2020est,Kawana:2021tde,Baker:2021nyl,Baker:2021sno}.

Vacuum energy stored in the false-vacuum remnants at the end of FOPTs can itself drive the formation of black holes, without the assistance of trapped matter or radiation~\cite{Flores:2024lng,Dent:2025bwo}. This possibility can be studied by directly following the spacetime evolution of false-vacuum domains, described by a set of boundary conditions, the Israel junction conditions, and determining when the enclosed vacuum energy satisfies the condition for Schwarzschild black hole formation~\cite{Blau:1986cw}. Unlike the conventional picture of PBH formation from primordial density fluctuations which relies on perturbations exceeding a numerically-determined critical overdensity~\cite{Carr:1975qj,Harada:2013epa,Escriva:2021aeh}, the collapse criterion here emerges directly from the gravitational dynamics of the infalling false-vacuum domain.
Once the domain reaches a turning point at which its own gravity overwhelms its expansion, subsequent collapse toward the formation of a black hole is inevitable. As a result, PBH production can occur without invoking matter asymmetries, finely tuned couplings, or large primordial fluctuations. This makes false-vacuum collapse a particularly robust and broadly applicable avenue for PBH production in FOPTs. Furthermore, while this mechanism is often associated with strongly supercooled FOPTs (that have a large separation in time between the onset of the transition and its completion), we show that false-vacuum collapse alone can efficiently produce PBHs across a broad region of parameter space, even in the absence of an extended dark sector.


Since PBH formation via vacuum collapse does not require trapping particles or satisfying overdensity thresholds, the currently established multi-messenger parameter space of GWs and PBH signatures (e.g. refs.~\cite{Baker:2021nyl,Marfatia:2021hcp,Xie:2023cwi,Gehrman:2023esa,Huang:2025hos,Huang:2022him,Lewicki:2024ghw,Arteaga:2024vde}) warrants reevaluation in this distinct framework. The GW spectrum and PBH abundance are both controlled by the underlying particle physics through the effective potential that governs the FOPT, and are therefore strongly correlated. Yet, the mapping between fundamental model parameters and the resulting GW-PBH phenomenology remains incomplete: it is not yet clear which regions of parameter space preferentially yield detectable GWs, detectable PBHs, or both simultaneously. A primary goal of this work is therefore to chart this parameter space for two representative classes of models: a generic polynomial effective potential and a classically conformal effective potential.

The paper is organized as follows. In \cref{sec:potentials} we first introduce an effective  polynomial potential before constructing a classically conformal U(1)$_{B-L}$ model and its finite-temperature effective potential. We then outline the machinery to describe the resulting phase transitions and cosmology in \cref{sec:PT_cosmology}. In \cref{sec:pheno} we discuss the GW signals and PBH formation mechanism. With all of this infrastructure in place, in \cref{sec:pspace} we perform scans over the parameter space of these models to show the breadth of possible GW and PBH signals, illustrating the multi-messenger regions of parameter space in \cref{sec:multimessenger}. Finally, in \cref{sec:conclusion} we conclude and discuss the impact of these findings.

\section{Effective Scalar Potentials at Nonzero Temperatures}~\label{sec:potentials}
We introduce two choices of scalar potentials described at finite temperature $T$ to illustrate their features and relevant model parameter space for FOPTs. We begin by considering a generic scalar potential $V$ described by quadratic, cubic, and quartic terms with coefficients that can be independently varied, including temperature-dependent terms. This first example serves as a flexible and largely model-independent benchmark for capturing a broad model parameter space for its phase transitions and resulting GW and PBH phenomenology.
We then turn to a model-specific realization of FOPTs within classically conformal U(1)$_{B-L}$ extensions of the SM. This scenario's effective potential $V_\mathrm{eff}$ incorporates one-loop-level contributions from both running couplings and nonzero temperatures. Included among the running couplings are the U(1)$_{B-L}$ gauge coupling, the quartic coupling of the symmetry-breaking scalar field, and Yukawa couplings between this scalar field and Majorana fields. Thermal corrections are brought on by VEV-dependent masses and Debye masses of bosons.

In any FOPT, the effective potential must develop a barrier separating the symmetric and broken phases (discussed in more detail in \cref{sec:PT_cosmology}). In the polynomial potential, this barrier is generated straightforwardly by a tree-level cubic term. By contrast, an analogous barrier emerges in the U(1)$_{B-L}$ effective potential through the interplay of renormalization-group flow and thermal corrections. Studying these two cases therefore allows us to distinguish generic features of GW and PBH production from those arising from nontrivial correlations between couplings and theoretical constraints of a realized model.


\subsection{A Polynomial Effective Potential}
\label{sec:generic_pt}
We can characterize a polynomial potential in terms of the modulus $\Phi$ of a complex scalar field at finite temperature with a common parameterization found in the literature,
\begin{equation}
\label{eq:generic_pot}
    V(\Phi, T) = D(T^2 - T_0^2) \Phi^2 - (AT + C)\Phi^3 + \frac{\lambda}{4} \Phi^4 \, ,
\end{equation}
for dimensionless coefficients $D$, $A$, and $\lambda$, and dimensionful coefficient $C$. This potential has a VEV at zero temperature of 
\begin{equation}
    v\equiv\langle\Phi\rangle_{T=0} = \dfrac{3C + \sqrt{9 C^2 + 8\lambda D T_0^2}}{2 \lambda} \, .
\end{equation}
The crossing point where the mass term changes sign is determined by the parameter $T_0$, which can be related directly to the VEV; equivalently, if we want to specify the VEV as an input, this determines $T_0$ as
\begin{equation}
    \label{eq:T0_generic}
    T_0 = \sqrt{\frac{\lambda  v^2-3 C v}{2D}} \, ,
\end{equation}
for $v = \phi_+$.
For the potential to be convex we require $\lambda > 0$ and for a vacuum expectation value (VEV) to appear at zero temperature we also require $T_0^2 > 0$ (and implicitly $D > 0$) imposing the constraint that $C < v \lambda / 3 \,$.
One can also determine the critical temperature $T_c$ such that the potential as a degenerate minima at both $\Phi = 0$ and $\Phi = \langle\Phi\rangle_{T\neq0}$,
\begin{equation}
    \label{eq:Tc_generic}
    T_c = \frac{1}{\lambda D - A} \bigg( A C + \sqrt{\lambda D (C^2 + (\lambda D - A^2) T_0^2)} \bigg) \, ,
\end{equation}
requiring that $A < \sqrt{\lambda D}$ for $T_c$ to be real. These positivity bounds and constraints define a parameter space to search for first-order phase transitions captured by potentials of this form, with the transition taking place between the temperatures $T_c$ and $T_0$.

We will employ this generic parameterization of a complex scalar field and its generation of a VEV in order to explore the parameter space supporting first-order phase transitions, and subsequent GW and PBH production. This serves as a benchmark comparison to the model-specific scenario of spontaneous breaking of conformal U(1)$_{B-L}$, discussed in the following subsection. Coleman-Weinberg terms can also be parameterized, although this form has the advantage of simple analytic and closed-form expressions for quantities of interest for describing phase transitions.

\subsection{A Classically Conformal Effective Potential}~\label{sec:B-L_pt}

The SM has an accidental global symmetry that conserves the difference between baryon and lepton number $B-L$. U(1)$_{B-L}$ models promote this to a gauged abelian symmetry and extend the SM as $\mathrm{SU}(3)_c \times \mathrm{SU}(2)_L \times \mathrm{U}(1)_Y \times \mathrm{U}(1)_{B-L}$, which may be further embedded within grand unified theories~\cite{Buchmuller:1991ce}.
Anomalies within the minimal U(1)$_{B-L}$-extended SM are canceled with the addition of three right-handed Majorana neutrinos $\nu_R^i$~\cite{Babu:1989ex} (where $i$ indexes the three generations), which have charges $Q_{B-L}^{\nu_R^i} = -1$ and a single internal degree of freedom $g_{\nu_R^i}=1$. These neutrinos may also provide a natural way to generate nonzero masses for the left-handed neutrinos in the SM via seesaw mechanisms~\cite{Gell-Mann:1979vob, Mohapatra:1979ia, Minkowski:1977sc, Yanagida:1979as}. Additional fields include a vector gauge boson $Z^\prime$ (with $Q_{B-L}^{Z^\prime} = 0$ and $g_{Z^\prime}=3$) required by gauge invariance, and a scalar boson $\Phi = (\phi + \mathrm{i} G)/\sqrt{2}$ (with $Q_{B-L}^{\Phi} = 2$ and $g_{\Phi}=1$) whose nonzero VEV spontaneously breaks the U(1)$_{B-L}$ symmetry. To be conformal at the classical level, the Lagrangian must not contain a mass term $-m_\Phi^2 \Phi^\dagger \Phi$.
In the following subsections we provide the symmetry-breaking effective potential at zero temperature before taking into account thermal effects. Further details on U(1)$_{B-L}$ modeling may be found in Refs.~\cite{Iso:2009ss, Iso:2009nw, Jinno:2016knw, Marzo:2018nov, Ellis:2020nnr}.

\subsubsection{Zero Temperature}
In addition to the Lagrangian terms of the SM, terms in its classically conformal U(1)$_{B-L}$ extension include
\begin{equation}
   \label{eq:B-L_lagrangian}
   \begin{split}
       \mathcal{L} &\supset \mathcal{L}_{\mathrm{kinetic}} + \mathcal{L}_{\mathrm{scalar}} + \mathcal{L}_{\mathrm{Yukawa}} \, ,
       \\
       \mathcal{L}_{\mathrm{kinetic}} &= (D_\mu \Phi)^\dagger (D^\mu \Phi) + \mathrm{i} (\nu_R^{i})^\dagger \, \overline{\sigma}^\mu D_\mu \nu_R^i \, ,
       \\
       \mathcal{L}_{\mathrm{scalar}} &= - \lambda (\Phi^\dagger \Phi)^2 \, ,
       \\
       \mathcal{L}_{\mathrm{Yukawa}} &= - \frac{1}{2} Y_i \Phi (\nu_R^{i})^\dagger \nu_R^i
       + \mathrm{h.c.} \, ,
   \end{split}
\end{equation}
where the covariant derivative for SM singlets reads $D_\mu \equiv \partial_\mu - \mathrm{i} g_{B-L} Z^\prime_\mu Q_{B-L}$ with spacetime derivatives $\partial_\mu$ and $B-L$ charge generator $Q_{B-L}$.
To focus on the U(1)$_{B-L}$ phase transition independently of electroweak effects, we neglect interactions between the SM and U(1)$_{B-L}$ sectors, including any kinetic mixing between the abelian groups. Furthermore, we assume a negligible Higgs-portal coupling, $\lambda^\prime \ll 1$, in the interaction term $-\lambda^\prime (\Phi^\dagger \Phi)(H^\dagger H)$ between $\Phi$ and the SM Higgs field $H$. We also neglect the Dirac neutrino Yukawa interaction $-y_{ij}(\nu_R^i)^\dagger H^\dagger l_L^j+ \mathrm{h.c.}$ involving electroweak lepton doublets $l$ of the SM, whose couplings are expected to be small in standard seesaw mechanisms with high U(1)$_{B-L}$-breaking scales.

The potential in $\mathcal{L}_\mathrm{scalar}$ admits radiative symmetry breaking~\cite{Coleman:1973jx} wherein running of couplings imparts a symmetry-breaking VEV $\vevMth$. This phenomenon is often studied through the renormalization group (RG)-improved effective potential for $\Phi$, given at the one-loop level by~\cite{Sher:1988mj, Meissner:2008uw}
\begin{gather}
    \label{eq:RG_eff_potential}
    V_0(\phi) = \frac{1}{4} \lambda(\tau) C(\tau)^4 \phi^4 \\
    C(\tau) \equiv \exp{ \left[ -\int_0^\tau \mathrm{d}\tau^\prime \gamma(\tau^\prime) \right] },
    \quad \gamma(\tau) \equiv \frac{1}{32 \pi^2} \left[ \sum_i Y_i^2 - a_2 g_{B-L}^2 \right],
    \quad a_2 = 24 \notag
\end{gather}
where $\phi = \sqrt{2} \, \mathrm{Re}(\Phi)$ and $\tau = \ln(\phi / \mu)$ with renormalization scale $\mu$. The RG equations are
\begin{align}
    \label{eq:RGEs}
    2\pi \diff{\alpha_{B-L}}{\tau} & = b \alpha_{B-L}^2, \\
    2\pi \diff{\alpha_\lambda}{\tau} & = a_1 \alpha_\lambda^2 + 8\pi \alpha_\lambda \gamma + a_3 \alpha_{B-L}^2 - \frac{1}{2} \sum_i \alpha_{Y_i}^2, \notag \\
    \pi \diff{\alpha_{Y_i}}{\tau} & = \alpha_{Y_i} \left( \frac{1}{2} \alpha_{Y_i} + \frac{1}{4} \sum_j \alpha_{Y_j} - 9 \alpha_{B-L} \right) \, , \notag
\end{align}

\begin{gather*}
    \alpha_{B-L} \equiv \frac{g_{B-L}^2}{4\pi}, \quad \alpha_\lambda \equiv \frac{\lambda}{4\pi}, \quad \alpha_{Y_i} \equiv \frac{Y_i^2}{4\pi} \, , \\
    b = 12, \quad a_1 = 10, \quad a_3 = 48 \, .
\end{gather*}
This analysis takes the renormalization scale to be the VEV of $\Phi$: $\mu = \vevMth$. The VEV (at zero temperature) is defined by the stationary condition, 
\begin{equation}
    \label{eq:stationary_condition}
    \diff{V_0}{\phi} \bigg\rvert_{\phi = \vevMth} = 0 \, ,
\end{equation}
which leads to the following relation between couplings evaluated at $\phi = \vevMth$:
\begin{equation}
    \label{eq:couplings_relation}
    \diff{\alpha_\lambda}{t} + 4 \alpha_\lambda (1-\gamma) = 0 \quad \quad (\tau=0) \, .
\end{equation}

The $\phi$-dependent masses are derived from Eqs.~\eqref{eq:B-L_lagrangian} with $\Phi = (\phi + \mathrm{i} G) / \sqrt{2}$ by taking second-order derivatives of the resulting tree-level potential at $G=0$. The $Z^\prime$ mass is found in $\mathcal{L}_{\mathrm{kinetic}}$, $\phi$ and Goldstone boson $G$ masses in $\mathcal{L}_{\mathrm{scalar}}$, and $\nu_R^i$ masses in $\mathcal{L}_{\mathrm{Yukawa}}$:\footnote{
While we neglect interactions between the SM Higgs $H$ and $\Phi$, we make note of two things regarding the $\phi$-dependent Higgs mass $m_H(\phi)^2 = \lambda^\prime(\tau) \phi^2$. First, our negligible $\lambda^\prime$ is naturally justified for any $\vevMth \gg 125$~GeV as $|\lambda^\prime(0)| = m_H^2 / \vevMth^2 \approx (125 \, \mathrm{GeV})^2 / \vevMth^2$.
Second, although we neglect coupling between $H$ and $\Phi$, many U(1)$_{B-L}$ studies consider $\lambda^\prime < 0$, as this leads to a negative $m_H^2$ thereby driving the electroweak phase transition. 
}
\begin{align}
\label{eq:phi-dependent masses}
    m_{Z^\prime}(\phi)^2 &= 4 g_{B-L}(\tau)^2 \phi^2 \, ,
    & m_\phi(\phi)^2 &= 3 \lambda(\tau) \phi^2 \, ,
    \\
    m_G(\phi)^2 &= \lambda(\tau) \phi^2 \, ,
    & m_{\nu_R^i}(\phi)^2 &= \frac{1}{2} Y_i(\tau)^2 \phi^2 \, . \notag
\end{align}
As seen later in \cref{sec:B-L_pt}, Yukawa couplings are found not to significantly influence FOPTs, and we thus simplify our analysis by considering a single generation of right-handed neutrino: $Y_i \rightarrow Y$.

\subsubsection{Nonzero Temperature}
When studying properties of a dense early universe, conventional quantum field theory should be supplemented with nonzero-temperature effects. The effective potential for this U(1)$_{B-L}$ model is given by a sum of the one-loop RG-improved effective potential $V_0$ and a nonzero-temperature potential $V_T$, also calculated to the one-loop level~\cite{Quiros:1999jp} under the Parwani resummation prescription:
\begin{gather}
    \label{eq:effective_potential}
    V_{\mathrm{eff}}(\phi, T) = V_0(\phi) + V_T(\phi, T) \, , \\
    V_T(\phi, T) = \frac{T^4}{2\pi^2} \sum_j g_j J_j \left( \frac{m_j(\phi)^2}{T^2} + \frac{\Pi_j(T)}{T^2} \right) \, , \notag
\end{gather}
where $g_j$ is the number of internal degrees of freedom for particle $j$. $J_j$ are the thermal integrals,
\begin{equation}
    \label{eq:thermal_integrals}
    J_j(x) = (-1)^{2s_j} \int_0^\infty dy \, y^2 \ln{\left[ 1 - (-1)^{2s_j} e^{-\sqrt{x + y^2}} \right]} \, ,
\end{equation}
with particle $j$ having spin $s_j$. The Debye masses $\Pi_j$ account for the contributions from daisy Feynman diagrams of bosons, and are given by
\begin{align}
    \label{eq:Debye_masses}
    \Pi_{Z^\prime}(T) &= 4 g_{B-L}(\tau)^2 T^2 \, , \\
    \Pi_{\Phi}(T) &= \frac{T^2}{24} \left[ 24 g_{B-L}(\tau)^2 + 8 \lambda(\tau) + \sum_i Y_i(\tau)^2 \right] \, , \notag \\
    \Pi_G(T) &= \Pi_\Phi(T) \notag \, .
\end{align}
It should be noted that the $\phi$-dependent and Debye masses for (distinct) antiparticles are identical to their particle counterparts and must be included in the sum of Eq.~\eqref{eq:effective_potential}. Due to the multiple energy scales ($\phi$ and $T$) within the system, there is some ambiguity in parameterizing the renormalization group flow~\cite{Einhorn:1983fc, Ford:1992mv, Bando:1992wy, Ford:1994dt, Ford:1996hd, Casas:1998cf, Steele:2014dsa, Chataignier:2018aud}. Possible choices may include $\tau = \log(\max[\phi, T]/\mu)$ or $\tau = \log(\sqrt{\phi^2 + T^2}/\mu)$, for example. We choose to parameterize with
\[ \tau \equiv \log{\left( \frac{\phi}{\mu} \right)} \, , \]
because no significant differences in phase transition parameters were found in these different choices of $\tau$.

\section{Phase Transition and Cosmology}~\label{sec:PT_cosmology}
We now briefly outline the machinery adopted to describe a FOPT including true-vacuum bubble nucleation rates, percolation time, and cosmic expansion dynamics in a universe with both radiation and vacuum energy density. We recommend ref.~\cite{Athron:2023xlk} for a thorough review.

\begin{figure}[h]
    \centering
    \includegraphics[width=0.65\textwidth]{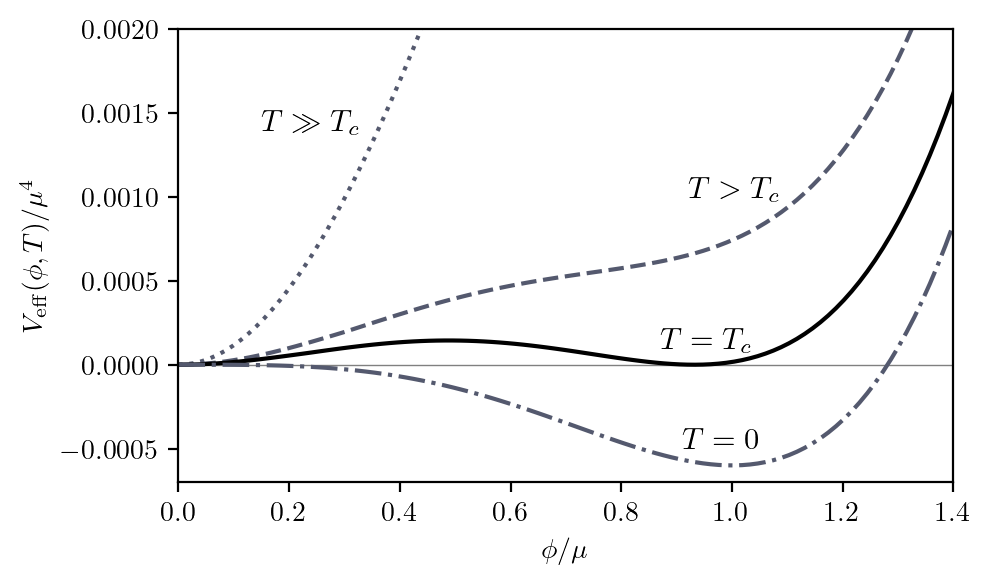}
    \caption{Effective potential progressing through a FOPT; at temperatures $T$ much greater than the critical temperature $T_c$, the expectation value of the field is stable at the origin, while at $T=T_c$ degenerate minima develop.}
    \label{fig:FOPT_B-L}
\end{figure}

Any particle model with symmetry breaking is accompanied by a cosmological phase transition, often marked by changes in field masses (e.g. Eqs.~\eqref{eq:phi-dependent masses}) as the symmetry-breaking scalar $\phi$ acquires its nonzero VEV. This transition is shown schematically in Fig.~\ref{fig:FOPT_B-L} and transpires as follows. Thermal corrections dominate $V_\mathrm{eff}$ at high temperatures and maintain a single global minimum for the potential at $\phi = 0$. As $T$ lowers, a new local minimum may form at some $\phi \ne 0$. For some critical temperature $T_c$, the new and original minima become degenerate. If the new minimum is located at $\phi = \phi_c$, this condition reads $V_\mathrm{eff}(0, T_c) = V_\mathrm{eff}(\phi_c, T_c)$. In a FOPT, a barrier in the potential separates the minima; that is $V_\mathrm{eff}(\phi, T_c) > V_\mathrm{eff}(0, T_c)$ for $\phi \in (0, \phi_c]$. At $T = T_c$ the $\phi$ field values at $0$ and $\phi_c$ are equally energetically favorable. However, for $T < T_c$ the off-origin minimum acquires energy densities below $V_\mathrm{eff}(0, T < T_c)$, and quantum tunneling away from $\phi = 0$ becomes more likely \cite{Coleman:1977py}. The physical picture is a universe initially in the $\phi = 0$ phase (also called the symmetric phase or false vacuum) at high temperatures. As the universe expands and cools, tunneling to a $\phi \ne 0$ phase (also called the broken phase or true vacuum) occurs stochastically throughout the universe. These domains of the new phase nucleate and expand as bubbles until the FOPT is complete.

The criteria for a FOPT outlined above are thus
\begin{equation}
    \label{eq:FOPT_criteria}
    V_\mathrm{eff}(0, T_c) = V_\mathrm{eff}(\phi_c, T_c), \quad \quad
    \diff{V_\mathrm{eff}}{\phi} \bigg\rvert_{\phi = \phi_c} = 0 \, ,
\end{equation}
which we use in our scans to find points in parameter space which yield such phase transitions.

\subsection{Bubble Nucleation}
At the high temperatures of the early universe, the probability per unit time per unit volume of true-vacuum nucleation is approximated semi-classically by~\cite{LINDE198137, Linde:1981zj, Enqvist:1991xw, Kierkla:2022odc}
\begin{equation}
    \label{eq:tunneling_probability_rate}
    \Gamma(T) \sim T^4 \left[ \frac{S_3(T)/T}{2\pi} \right]^{3/2} e^{-S_3(T)/T} \, ,
\end{equation}
where $S_3$ is the three-dimensional euclidean action,
\begin{equation}
    \label{eq:S_3_euclidean_action}
    S_3(T) = 4\pi \int_0^\infty \mathrm{d}r \, r^2 \left[ \frac{1}{2} \left( \diff{\phi_b}{r} \right)^2 + V_\mathrm{eff}(\phi_b(r), T) \right] \, ,
\end{equation}
with $r$ the radial spatial coordinate and $\phi_b(r)$ the scalar field value at a distance $r$ from the bubble's center~\cite{Linde:1977mm}. $\phi_b$ is the SO(3)-symmetric bounce solution satisfying
\begin{equation}
    \label{eq:bounce_solution}
    \begin{split}
        \diff[2]{\phi_b}{r} + \frac{2}{r} \diff{\phi_b}{r} = \diff{V_\mathrm{eff}(\phi, T)}{\phi} \bigg\rvert_{\phi = \phi_b} \, , \\
        \lim\limits_{r \rightarrow \infty} \phi_b(r) = \phi_\mathrm{true} \, , \quad \quad
        \diff{\phi_b}{r} \bigg\rvert_{r=0} = 0 \, ,
    \end{split}
\end{equation}
where $\phi_\mathrm{true}$ is the field value at the global minimum of the potential (the VEV) for a given temperature. We employ the Python package \texttt{ELENA}~\cite{Costa:2025pew} to calculate $S_3$ utilizing the tunneling potential method~\cite{Espinosa:2018hue}. To characterize when a given FOPT occurs, we use the percolation temperature $T_p$, also calculated with \texttt{ELENA}. This is the temperature at which domains of the true vacuum begin to form a connected cluster spanning the universe. Numerical studies~\cite{10.1063/1.1338506, LIN2018299, LI2020112815} find the onset of percolation to occur when about 29\% of the universe's 3-volume is in \textit{true} vacuum. Cosmological studies tend to use instead the fraction of the universe in \textit{false} vacuum $P_f$, in which case percolation temperature is defined by
\begin{equation}
    \label{eq:percolation_temperature}
    P_f(T_p) = 0.71 \, .
\end{equation}
We find that the sensitivity to phase transition quantities and observables in response to small variation in this threshold is very weak. The multi-messenger parameter space we identify later in \cref{sec:multimessenger} is therefore robust against this choice, and we adopt $P_f(T_p) = 0.71$ throughout.

The fractional volume in false vacuum is given at any cosmic time $t$ by~\cite{Guth:1979bh, Guth:1981uk}
\begin{equation}
    \label{eq:false_vacuum_fraction_Pf}
    P_f(t) = \exp{\left[ -\int_{t_0}^t \mathrm{d}t^\prime \, \Gamma(t^\prime) \frac{a(t^\prime)^3}{a(t)^3} V(t^\prime, t) \right]} \, ,
\end{equation}
where $t_0$ is the time at which the FOPT begins (chosen here to be when the critical temperature $T_c$ occurs) and $a(t)$ is the scale factor of a Friedmann-Lema\^{\i}tre-Robertson-Walker spacetime manifold. The physical 3-volume $V(t^\prime, t)$ of a false-vacuum domain nucleated at $t^\prime$ is assumed spherical,
\begin{align}
    \label{eq:bubble_radius}
    V(t^\prime, t) &= \frac{4\pi}{3} R(t^\prime, t)^3 \, , \\
    R(t^\prime, t) &= \frac{a(t)}{a(t^\prime)} R_0(t^\prime) + \int_{t^\prime}^t \mathrm{d}t^{\prime \prime} \, \frac{a(t)}{a(t^{\prime\prime})} v_w(t^{\prime\prime}) \, ,
\end{align}
where $R(t^\prime, t)$ is the radius at $t$ of a bubble nucleated at $t^\prime$, and $R_0(t^\prime)$ gives the initial radius of this bubble---which we take to be negligible. While the wall speed $v_w(t)$ relative to the background plasma varies and ultimately depends on the particle-scale interactions at the bubble wall~\cite{Ellis:2019oqb,Kierkla:2022odc}, we assume $v_w = 1$ for this analysis.

\subsection{Cosmology}
Standard cosmology in a flat FLRW spacetime is used to calculate the ratios of scale factors in preceding equations:
\begin{equation}
    \label{eq:scale_factor_ratio}
    \frac{a(t_1)}{a(t_2)} = \exp\left[ \int_{t_2}^{t_1} \mathrm{d} t^\prime H(t^\prime) \right] \, ,
    \qquad
    H^2 = \frac{8\pi G}{3} (\rho_V + \rho_R) \, ,
\end{equation}
where $H$ is the Hubble parameter and $G = 6.708 \times 10^{-39} \, \mathrm{GeV}^{-2}$ is the gravitational constant. The energy densities in vacuum and radiation are
\begin{equation}
    \label{eq:energy_densities}
    \rho_V(T) = \Delta V_{\rm eff}(T) - T \diffp{\Delta V_{\rm eff}(T)}{T} \, ,
    \qquad
    \rho_R(T) = \frac{\pi^2}{30} g_R(T) T^4 \, ,
\end{equation}
with $\Delta V_{\rm eff}(T) := V_{\rm eff}(\phi_{\rm false}, T) - V_{\rm eff}(\phi_{\rm true}, T)$ and $g_R$ gives the number of degrees of freedom in radiation (see ref.~\cite{Borsanyi:2016ksw} for the temperature dependence of $g_R$ in the SM). We let $\phi_{\rm false} = 0$ and apply a (temperature-dependent) shift to the potential to keep $V_{\rm eff}(\phi_{\rm false}, T) = 0 \,\, \forall \, T$.

To convert between time and temperature, we assume entropy is conserved during universal expansion, which leads to
\begin{equation}
    \label{eq:temperature_time_derivative}
    \diff{T}{t} = -3H \frac{s}{\mathrm{d} s / \mathrm{d} T} =
    3H\frac{\diffp{V_{\rm eff}(\phi_{\rm false})}{T} - \frac{1}{3} \diffp{\rho_{R}}{T}}{\diffp[2]{V_{\rm eff}(\phi_{\rm false})}{T} - \frac{1}{3} \diffp[2]{\rho_{R}}{T}} \, ,
\end{equation}
noting that entropy density $s$ is given by the negative temperature derivative of free energy density. Throughout this work, Eq.~\eqref{eq:temperature_time_derivative} is used to convert all integrals over time into integrals over temperature.

\section{Gravitational Wave and Primordial Black Hole Phenomenology}
\label{sec:pheno}


\subsection{Gravitational Waves}
Particularly strong FOPTs could displace enough energy to produce a stochastic GW background detectable by future experiments which aim to measure the amplitudes and frequencies of GWs. FOPTs that release more latent heat often produce GWs of greater amplitudes which are easier to detect. Given the stochastic nature of FOPTs, their GW signal is expected to yield a wide range of frequencies peaked around some frequency heavily influenced by the characteristic length scale of the phase transition. These two properties of a FOPT, its latent heat and length scale, are often characterized by $\alpha$ and $R_{\rm sep}$:
\begin{equation}
    \label{eq:alpha_Rsep}
    \begin{split}
        \alpha &= \frac{\Delta \Bar{\theta}(T_p)}{3 w(\phi_{\rm false}, T_p) / 4} \, ,
        \\
        R_{\rm sep} &= \left[ \int_{t_0}^{t_p} \mathrm{d}t \, \Gamma(t) P_f(t) \frac{a(t)^3}{a(t_p)^3} \right]^{-1/3}
        \, .
    \end{split}
\end{equation}
The length scale $R_{\rm sep}$ is the mean separation between true-vacuum domains at percolation time $t_p$.

In defining $\alpha$, enthalpy density is given by $w(\phi, T) = -T \diffp{V_{\rm eff}(\phi, T)}{T}$. We also use $\Delta \Bar{\theta}(T)$ which gives the difference between false and true vacuum of the psuedotrace anomaly of the energy-momentum tensor~\cite{Giese:2020rtr}, and simplifies to
\begin{equation}
    \label{eq:psuedotrace_anomaly}
    \Delta \Bar{\theta}(T) := \frac{1}{4} \left(\rho_V(T) + \frac{\Delta V_{\rm eff}(T)}{c_{s}(T)^2}\right) \, ,
\end{equation}
where $c_{s}$ is the speed of sound in the universal plasma in true vacuum. This psuedotrace anomaly generalizes the trace anomaly for models with sound speeds not generally given by $c_{s}^2 = 1/3$, but rather
\begin{equation}
    \label{eq:sound_speed}
    c_{s}(T)^2 = \frac{\partial V_{\rm eff} / \partial T}{T \, \partial^2 V_{\rm eff} / \partial T^2} \bigg\rvert_{\phi = \phi_{\rm true}} \, ,
\end{equation}
which assumes the temperatures across domain walls in the FOPT are equal. 

In a FOPT, the three dominant contributions to a GW background come from sound waves in the plasma (sw), collisions of true vacuum bubbles (col), and magnetohydrodynamic turbulence in the plasma (turb). Following ref.~\cite{Caprini:2015zlo}, we approximate the total GW background to be a linear combination of these stochastic sources: $\rho_\mathrm{GW} = \rho_\mathrm{sw} + \rho_\mathrm{col} + \rho_\mathrm{turb}$
with $\rho$ the corresponding energy density. GW signal spectra are usually presented in terms of the logarithmic derivative~\cite{Maggiore:1999vm, PhysRevD.59.102001}
\begin{equation}
    \label{eq:total_GW_spectrum}
    \Omega_\mathrm{GW} \equiv \frac{1}{\rho_c(t_0)} \frac{\mathrm{d}\rho_\mathrm{GW}}{\mathrm{d}\ln(f_\mathrm{GW})} \, ,
\end{equation}
where $f_\mathrm{GW}$ is the GW frequency and $\rho_{c}(t_0) = \frac{3 H_0^2}{8 \pi G} = 8.095895 \times 10^{-47} \, h^2 \, \text{ GeV}^4$ gives today's critical energy density. Numerical fits for each contribution are obtained from hydrodynamical simulations. The GW module within \texttt{ELENA} is used to calculate these spectra, which employs fits from ref.~\cite{Ellis:2019oqb}. We update the peak frequencies from sound waves to match findings from ref.~\cite{Caprini:2019egz}. The complete set of parameterized fits adopted in this work can be found in Appendix~\ref{app:gwfits}.

\subsection{Primordial Black Hole Formation}
We now turn to the possibility of PBH formation from the direct collapse of false vacuum domains towards the end of the phase transition. A common method to treat the formation of PBHs uses the notion of the energy density contrast related to curvature perturbations and the critical overdensity criterion to determine formation~\cite{Carr:1975qj,Harada:2013epa,Escriva:2021aeh}. This method is not germane to post-inflationary PBH formation, since the critical overdensity is not a geometry-agnostic estimator of the collapse condition~\cite{Flores:2024lng}. Instead, we can model the energy stored in the false-vacuum bubbles of the scalar field to drive the collapse to PBHs directly from the Schwarzschild criterion. To describe this, we adopt the formalism originally developed in ref.~\cite{Blau:1986cw}, which uses the language of the Israel junction condition, a continuity relation between the interior and exterior spacetimes of the bubble, in order to derive an equation of motion for the radius of the false-vacuum bubble. This and similar approaches to the false vacuum collapse were recently explored in refs.~\cite{Lewicki:2023ioy, Flores:2024lng,Murai:2025hse,Hashino:2025fse}, especially in the context of supercooled transitions, though the precise strategy may depend on the background spacetime metric and radiation/vacuum energy content at the time of the transition. We will repeat some of the details here for the convenience of the reader.

Namely, we consider the mass $M$ contained in a false vacuum patch separated by a spherical domain wall with radius $r$;
\begin{equation}
\label{eq:PBH_mass_Misner_Sharp}
    M = \frac{4}{3}\pi \rho_V r^3 - 8\pi^2 G \sigma^2 r^3 + 4\pi\sigma r^2 \bigg[ 1 - \bigg(\frac{8\pi G}{3} \rho_V \bigg) r^2 + \bigg(\frac{dr}{d\tau}\bigg)^2 \bigg]^{1/2} \, .
\end{equation}
Here, $\sigma$ is the wall tension and $\rho_V$ is the vacuum energy density taken as the potential difference $\Delta V$ between the true and false vacuum at the time of the collapse, i.e., at $T_{\rm PT}$. It is important to note that Eq.~\eqref{eq:PBH_mass_Misner_Sharp} holds for regions with only (true and false) vacuum components. To use this formalism for our analysis, we ensure that SM radiation negligibly contributes to the local energy density by only considering models with wall radii that are much smaller than the ``radiation Hubble'' radii at the time of PBH formation:
\begin{equation}
\label{eq:BGG_condition}
    r \ll \left(\frac{8\pi G}{3} \rho_R \right)^{-1/2}
\end{equation}
which maintains the validity of the assumption of vacuum domination~\cite{Blau:1986cw} (Appendix~\ref{app:BGG_validity} gives further details). The differential equation~\eqref{eq:PBH_mass_Misner_Sharp} in the wall position $r$ can be reframed into an equation of motion of a particle moving in the presence of a potential,
\begin{equation}
    \bigg( \dfrac{d z}{d\tau^\prime}\bigg)^2 + U(z) = E \, ,
\end{equation}
where $E$ and $U(z)$ are defined in ref.~\cite{Flores:2024lng} and depend on $M$, $\rho_V$, and $\sigma$, with the dimensionless parameter $z$ defined as
\begin{equation}
    z(r) = \frac{1}{2 G M}\bigg( \frac{1}{3}G\rho_V + \frac{1}{4}G^2 \sigma^2 \bigg)^{1/3} r \,.
\end{equation}
The mass $M$ contained in the false vacuum bubble is taken as a constant and dominated by the volume term $\frac{4}{3}\pi \rho_V r^3$.

We can estimate the initial formation size of the false-vacuum domain with a choice of its characteristic size; one good choice is the mean radius of this patch $\bar{R}$,
\begin{equation}
    \bar{R} = \frac{1}{n} \int_{t_c}^t \mathrm{d}t^\prime \, \Gamma_f(t^\prime) \left[1 - P_f(t^\prime)\right] \bigg[ \frac{a(t^\prime)}{a(t)}\bigg]^3 R(t^\prime, t) \, .
\end{equation}
Note the differences in this equation from its usual treatment as the mean radius of \textit{true}-vacuum domains~\cite{Athron:2023xlk}. Instead, we adopt it to describe the nucleation rate for the false vacuum domains $\Gamma_f$ as in ref.~\cite{Lu:2022paj} using a geometric approach;
\begin{equation}
\label{eq:lu_rate}
    \Gamma_f(t) \equiv 32 \pi^4 v_w^9 P_f(t) \prod_{i=0; t_0 \equiv t}^3 \int_{t_c}^{t_i} \mathrm{d}t_{t+1} \bigg( \int_{t_{i+1}}^t \mathrm{d}t^\prime \frac{a(t_{i+1})}{a(t^\prime)} \bigg)^2 \Gamma(t_{i+1}) \frac{a(t_{i+1})}{a(t)} \, ,
\end{equation}
where $\Gamma$ is the bubble nucleation rate.
The number density $n$ is also defined;
\begin{equation}
    n_{\rm PBH}  = \int_{t_c}^t \mathrm{d}t^\prime \Gamma_f(t^\prime) \left[1 - P_f(t^\prime)\right] \bigg[ \frac{a(t^\prime)}{a(t)}\bigg]^3 \, .
\end{equation}
To implement these modified computations of the false vacuum nucleation rate and density, we have extended a local version of the \texttt{ELENA} package.

For successful collapse to take place, we impose the same constraints as in ref.~\cite{Flores:2024lng}, namely that the bubble nucleation rate and the Hubble time are both slower than the collapse time,
\begin{equation}
    \tau^\prime = \int_{z_s}^{z^*} \dfrac{\mathrm{d}z^\prime}{\sqrt{E - U(z)}} \, .
\end{equation}
The motion of the vacuum domain wall is integrated from $z^* = z(R^*)$ until the Schwarzschild limit $z_s$ corresponding to $z_s = z(2 G M)$. A PBH is considered to be formed if the initial size $z^*$ is beyond the ``turning point'' limit, beyond the potential barrier $U(z)$ and if the collapse time satisfies the above conditions.
If the collapse is satisfied for a given phase transition, $\bar{R}$ defined above directly allows us to estimate the PBH mass $M$ at the formation time. Assuming a monochromatic mass spectrum formed this way, the abundance can then be computed.

The false-vacuum collapse mechanism considered above is not the only possible mechanism for PBH formation during FOPTs. Another mechanism, like the one considered in ref.~\cite{Baker:2021nyl}, relies on the buildup of an overdensity of fermions inside the shrinking false-vacuum domains. In the case of the U(1)$_{B-L}$ model, heavy right-handed neutrinos can play the role of trapped particles as long as their masses $m_{\nu_R^i} = Y_i \langle\Phi\rangle/\sqrt{2}$ are sufficiently larger than $T_p$ to kinematically block them from penetrating domain walls, crossing from the unbroken to the broken phase. To achieve this, the mechanism may require some fine-tuning since the larger Yukawa couplings required also exhibit a more constrained space for possible first-order transitions (see Fig.~\ref{fig:B-L_FOPT_scan}). Furthermore, $\nu_R^i - {\nu_R^i}^\dagger$ annihilation through their Yukawa interaction can stop the overdense population from becoming compact enough to collapse into a black hole as the false-vacuum domain contracts. To maintain enough particles within this region, there are typically two scenarios considered. The first utilizes a number asymmetry between $\nu_R^i$ and ${\nu_R^i}^\dagger$ so that one species survives any amount of annihilation~\cite{Kawana:2021tde, Hong:2020est}. The second scenario realizes a suppressed annihilation rate through small Yukawa couplings $Y_i$. As shown in Sec.~\ref{sec:B-L_PT_scan}, Yukawa couplings are found to be inconsequential to FOPTs, and may be as small as needed, provided they are large enough to maintain the trapping condition $m_{\nu_R^i} \gg T_p$. Since we don't presume any number asymmetry in the right-handed neutrinos and the Yukawa couplings would have to be finely tuned in the second trapping scenario, we maintain the assumption that PBH formation is driven by vacuum energy alone.

\subsection{Primordial Black Hole Abundance}
 The resulting abundance of PBHs today is expressed relative to today's dark matter abundance, $f_\mathrm{PBH} = \Omega_\mathrm{PBH}/\Omega_\mathrm{DM}$, where we take the dark matter abundance at its central observed value, $\Omega_\mathrm{DM} h^2 = 0.1200 \pm 0.0012$~\cite{Planck:2018vyg} where $h$ is determined by the measurement of today's Hubble parameter $H_0 = 100 \, h \, \text{ km} \text{ s}^{-1} \text{ Mpc}^{-1}$. Generally, $f_\mathrm{PBH}$ will depend on the range of PBH masses making up $\Omega_\mathrm{PBH}$. At the time of their formation $t_f$, the energy density in PBHs is
\begin{equation}
    \label{eq:rho_PBH_formation_PT}
    \rho_\mathrm{PBH}(t_f) = \int_{M_\mathrm{min}}^{M_\mathrm{max}} \mathrm{d} M \, M \diff{n_\mathrm{PBH}}{M}
    = M n_{\rm PBH} \, ,
\end{equation}
where $[M_\mathrm{min}, M_\mathrm{max}]$ is the window of initial PBH masses, and we have assumed a monochromatic PBH mass distribution for simplicity.

This initial PBH density must be extrapolated to today's expected value to find $f_\mathrm{PBH}$:
\begin{equation}
    \label{eq:f_PBH_fraction_today}
    \begin{split}
        f_\mathrm{PBH} &= \frac{\Omega_\mathrm{PBH}}{\Omega_\mathrm{DM}} = \frac{\rho_\mathrm{PBH}(t_0)}{\Omega_\mathrm{DM} \rho_c(t_0)}
        = \frac{M n_{\rm PBH}(t_0)}{\Omega_{\rm DM} \rho_c(t_0)}
        = \frac{M n_{\rm PBH}(t_f)}{\Omega_{\rm DM} \rho_c(t_0) } \left[ \frac{a(t_f)}{a(t_0)} \right]^3
        \approx \frac{M n_{\rm PBH}(t_f)}{\Omega_{\rm DM} \rho_c(t_0)} \frac{s(t_0)}{s(T_{\rm reh})} \, ,
    \end{split}
\end{equation}
where today's entropy density is
$s(t_0) = 2.2215\times 10^{-38} \, \text{ GeV}^4$~\cite{ParticleDataGroup:2024cfk}, and we've approximated matter and entropy densities to scale similarly: $[a(t) / a(t^\prime)]^{-3} \approx s(t) / s(t^\prime)$, with equality holding only in adiabatic expansion. Significant reheating from frictional interactions between vacuum domain walls and the surrounding plasma can inject entropy into the system, and we use the reheating temperature $T_{\rm reh} \approx (1+\alpha)^{1/4} T_p$~\cite{Leitao:2015fmj} to help account for this. PBHs evaporating through their Hawking radiation~\cite{Hawking:1975vcx} is another phenomenon expected to dilute their mass density, and therefore $f_{\rm PBH}$. However, significant changes in mass due to this evaporation are expected only for PBHs with masses near roughly $10^{15}$~g (see ref.~\cite{Carr:2020gox}). We find the most phenomenologically interesting points in our models to yield PBH masses well above this scale (see Fig.~\ref{fig:B-L_f_PBH_VEVs}), and thus ignore evaporation effects.

Finally, we must specify a time to characterize when PBHs form. Just as percolation of the \textit{true} vacuum signals the onset of a FOPT, we use the percolation of the \textit{false} vacuum to demarcate PBH formation, defined through
\begin{equation}
    \label{eq:false_vacuum_percolation}
    P_f(T_f) = 0.29 \, .
\end{equation}
This provides a more reasonable characterization than true-vacuum percolation because below $T_f$ the connected cluster of false-vacuum domains begins to fragment and form isolated pockets potentially accommodating PBH formation. We find the impact on PBH observables is robust to changes in this convention; $\mathcal{O}(1)$ variation in $P_f$ results in shifts to observables like $\fpbh$ at the percent level.
We now have definitions to rewrite Eq.~\eqref{eq:BGG_condition}, which provides a necessary condition for applying the vacuum-collapse mechanism of PBH production~\cite{Blau:1986cw} used throughout this analysis:
\begin{equation}
    \overline{R}(T_f) \ll \left(\frac{8\pi G}{3} \rho_R(T_f) \right)^{-1/2} \, .
\end{equation}
This is found to be easily satisfied in the models considered below.

\section{Scans Over the Parameter Space}~\label{sec:pspace}
In this section we describe the procedure of scanning through the model parameter space for the case of the polynomial potential described in \cref{sec:generic_pt} and for the U(1)$_{B-L}$ potential described in \cref{sec:B-L_pt}. At each point in model parameter space we compute the resulting GW signal and PBH abundance that may result using the tools presented in \cref{sec:pheno}.

\subsection{Phase Transitions of a Generic Polynomial Potential}
We first examine the phase transitions corresponding to the polynomial potential model in Eq.~\eqref{eq:generic_pot}. We start by specifying a VEV $\vevMth$ and determine $T_0$ and $T_c$ with the positivity constraints described below Eq.~\eqref{eq:T0_generic} and Eq.~\eqref{eq:Tc_generic}. Then, we proceed to draw random variates of the parameters in the following ranges according to uniform distributions $U$,
\begin{align}
    \log_{10}\lambda &\in [-3, 0] \, , \nonumber \\
    A &\in [0.1 A_{\rm max}, A_{\rm max}] \, , \nonumber \\
    C &\in [0.1 C_{\rm max}, C_{\rm max}] \, , \nonumber  \\
    D &= 0.1 \, ,
\end{align}
where $C_{\rm max} = v \lambda / 3$ and $A_{\rm max} = \sqrt{\lambda D}$ in accordance with real positivity of $T_0$ and $T_c$ as discussed in \cref{sec:generic_pt}. The various temperatures (percolation and nucleation) are derived using \texttt{ELENA} along with other key quantities relevant to the calculation of the GW and PBH spectra described above. These ranges have been chosen to approximately constrain us to the phenomenologically relevant portion of parameter space such that the strength of the transition $\alpha \lesssim 10$. We perform a scan over model parameters in this way for benchmark VEVs from \vev$=1$ MeV up to 10 TeV.

\subsection{Phase Transitions of a Classically Conformal Effective Potential} \label{sec:B-L_PT_scan}
While assuming negligible $\lambda^\prime$ and a single Yukawa coupling, the free parameters of our classically conformal U(1)$_{B-L}$ model include three couplings \{$g_{B-L}$, $\lambda$, $Y$\}, and the VEV $\vevMth$. However, the condition of Eq.~\eqref{eq:couplings_relation} allows for the determination of one coupling from the other two at $\tau=0$. With the RG evolution of Eqs.~\eqref{eq:RGEs}, there are then two free couplings in this model which we choose to be $\alpha_{B-L}(0)$ and $\alpha_{Y}(0)$. We take all free parameters to be real and the couplings perturbative ($\alpha_{B-L}(0) < 1, \, \alpha_{Y}(0) < 1$).
In searching for phase transitions, we numerically scan the two-dimensional parameter space $\{\alpha_{B-L}(0), \, \alpha_{Y}(0)\}$ to find points satisfying the FOPT criteria Eqs.~\eqref{eq:FOPT_criteria}.

\begin{figure}[ht!]
    \centering
    \includegraphics[scale=0.2]{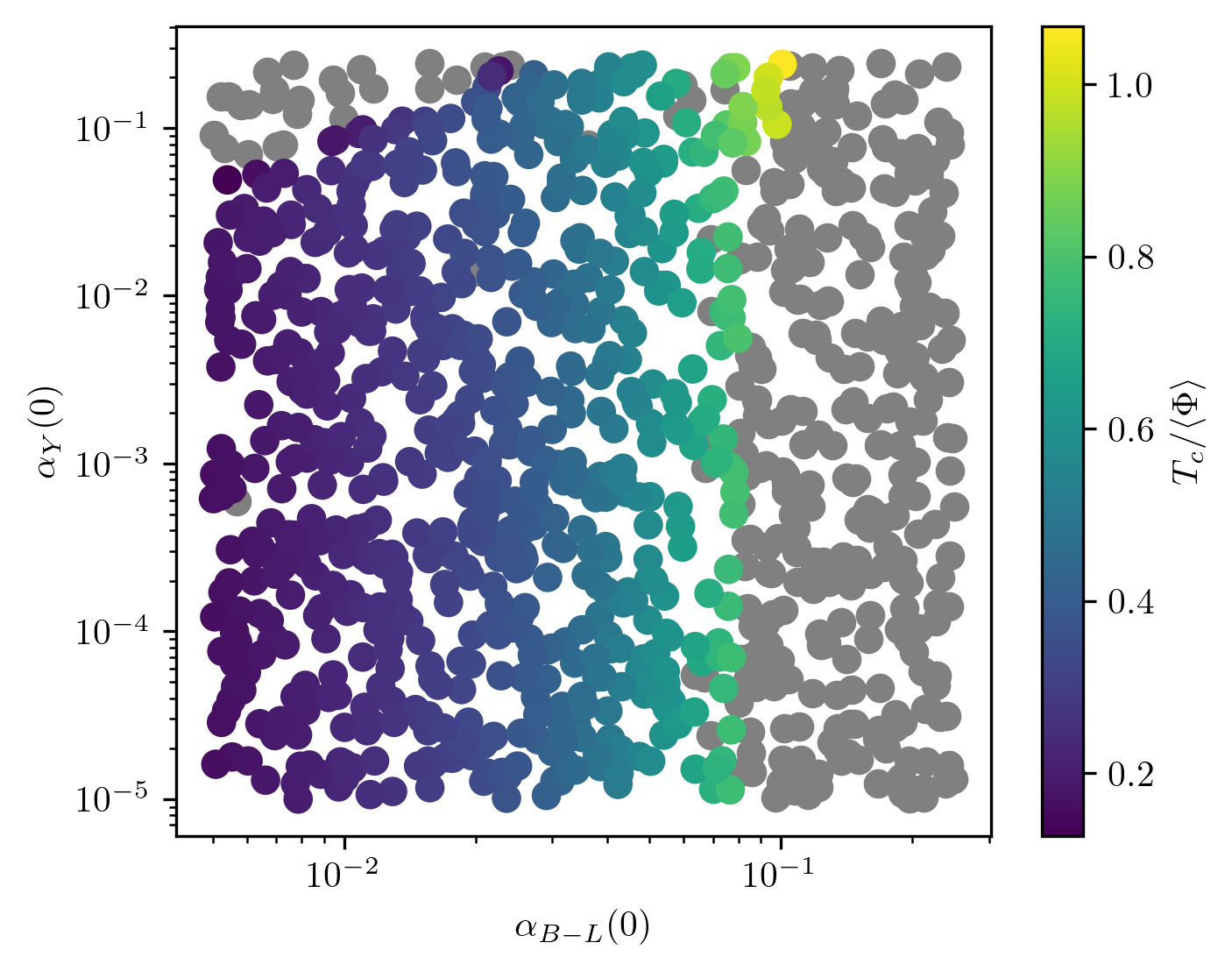}
    \caption{Classically conformal U(1)$_{B-L}$ parameter space with colored points showing couplings yielding successful FOPTs. The gray points were tested and found not to exhibit FOPTs.}
    \label{fig:B-L_FOPT_scan}
\end{figure}

Fig.~\ref{fig:B-L_FOPT_scan} shows a representative sample of classically conformal U(1)$_{B-L}$ models tested for FOPTs, with colored points indicating models found to produce FOPTs. When scanning over the ranges $\alpha_{B-L} \in [5 \times 10^{-3}, 0.25]$ and $\alpha_{Y} \in [10^{-5}, 0.3]$, it is found that FOPT properties are mostly agnostic to the Yukawa coupling between $\Phi$ and $\nu_R$. The same cannot be said about the U(1)$_{B-L}$ gauge coupling, as Fig.~\ref{fig:B-L_FOPT_scan} already shows its effect on critical temperatures. Additionally, decreasing $\alpha_{B-L}$ tends to make shallower the true-vacuum well in the effective potential; that is, for temperatures below $T_c$, $\Delta V_\mathrm{eff}(T)$ tends to be smaller in models with smaller gauge couplings. This weakens true-vacuum nucleation rates in such models, and we explore below the consequences on GW signals and PBH abundance in these delayed FOPTs. While no FOPTs were found in the gray points of Fig.~\ref{fig:B-L_FOPT_scan}, it is not immediately clear whether these regions are completely barren due to numerical ambiguities. In the region of parameter space with $\alpha_Y(0) \gtrsim 10 \alpha_{B-L}(0)$ (upper-left), sporadic RG flow moving between negative and positive couplings leads to numerical instabilities. For points with $\alpha_{B-L}(0) \gtrsim 0.1$, we run into Landau poles in the quartic coupling.

In the minimal U(1)$_{B-L}$ models considered here, the gauge coupling and gauge field's $\phi$-dependent mass are related in Eqs.~\eqref{eq:phi-dependent masses} as $m_{Z^\prime}(\phi) = 2 g_{B-L} \phi$. Existing beam-dump and neutrino-scattering searches at particle accelerators place strong limits on light $Z^\prime$ gauge bosons which exclude much of the sub-GeV parameter space at the sizeable gauge couplings scanned in this work~\cite{Graham:2021ggy}. Our U(1)$_{B-L}$ results are therefore best viewed as complementary to accelerator constraints: when translated into the same parameter space, considerations from gravitational lensing and PBH overproduction can further restrict disfavored models. 



\subsection{Predictions: Gravitational Wave Signals}
\subsubsection{Polynomial Effective Potential}
\begin{figure}[ht!]
    \centering
    \includegraphics[width=0.8\textwidth]{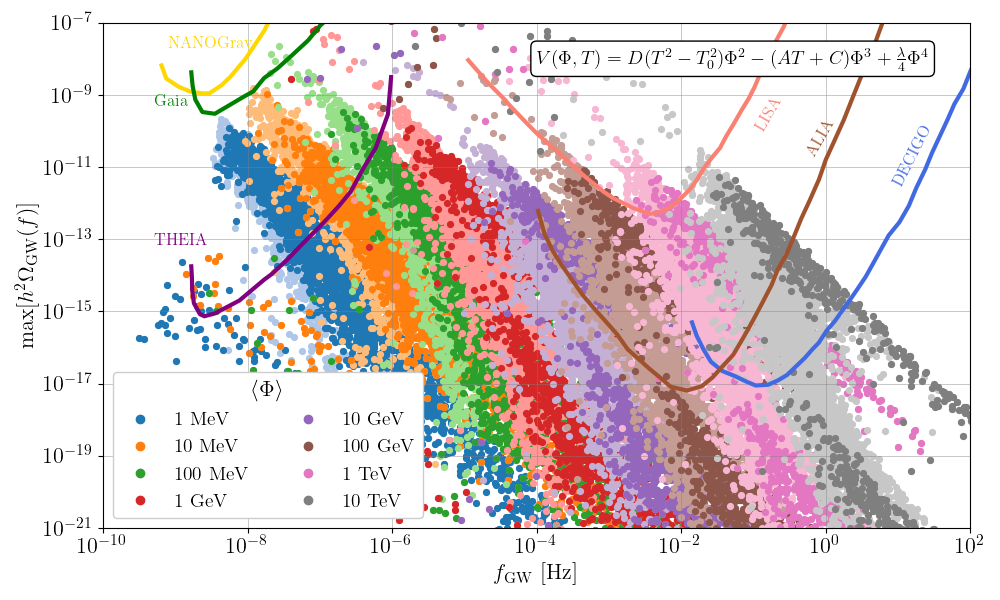}
    \caption{Gravitational wave abundances taken at the peak strain ($h^2 \Omega_{\rm GW}$) as a function of peak frequency for a scan of the parameters for the generic scalar field potential. The darker shaded points represent those that do not overproduce PBHs, e.g. $f_{\rm PBH} < 1$, while the lighter shaded points coincide with overabundant PBH production.}
\label{fig:gw_generic}
\end{figure}

We show the resulting distribution of GW strains by scanning over the generic polynomial potential parameters, as described above, by capturing the peak strain, $\max [h^2 \Omega_{\rm GW}]$, (dominated by the sound wave contribution, as collisional and turbulence contributions are smaller for the parameter space of interest, see \cref{app:gwfits}) for each successful first-order phase transition in Fig.~\ref{fig:gw_generic}. These strain peaks are compared with projected sensitivity reaches of future GW detectors like NANOGrav~\cite{NANOGrav:2020bcs, Lazio:2017fos}, Gaia~\cite{brown2018gaia}, THEIA~\cite{2018FrASS...5...11V}, LISA~\cite{Caprini:2019egz, amaro2017laser, Robson:2018ifk}, ALIA~\cite{Gong:2014mca}, and DECIGO~\cite{Kawamura:2020pcg, Kawamura:2006up}. See also refs.~\cite{EPTA:2011kjn,Ruan:2018tsw,TianQin:2015yph,Corbin:2005ny, Yagi:2011wg,Shoemaker:2019bqt, LIGOScientific:2014pky,A+:LIGO,Punturo:2010zz,LIGOScientific:2016wof} for other proposed experimental projections not shown here.

For each VEV in Fig.~\ref{fig:gw_generic}, the lighter VEVs occupy lower GW frequencies, approaching nHz scales at \vev$=1$ MeV, due to the correlation with low $\beta$ at small \vev, as expected. Within each scan for a fixed VEV, the extent over GW frequencies and strains is strongly correlated to the size of the cubic couplings ($A,C$) and the quadratic coupling ($\lambda$); smaller quadratic couplings for a fixed \vev have the effect of deepening the potential well $\Delta V$, thereby increasing the strength of the transition $\alpha$ and the resulting GW strain. Lowering the size of the cubic couplings has a similar effect under fixed \vev, due to the minus sign next to the cubic terms in the definition of the potential in Eq.~\eqref{eq:generic_pot}.

In Fig.~\ref{fig:gw_generic}, we also indicate which model points led to underabundant ($\fpbh < 1$, darker shading) and overabundant ($\fpbh > 1$, lighter shading) production of PBHs. As we will discuss in more detail in \cref{subsec_PBH_results} and \cref{sec:multimessenger}, the high efficiency of PBH production from large vacuum energy at higher VEVs leads to parametrically overabundant PBH production.

\subsubsection{Classically Conformal Effective Potential}
\begin{figure}[h!]
    \centering
    \includegraphics[width=0.8\textwidth]{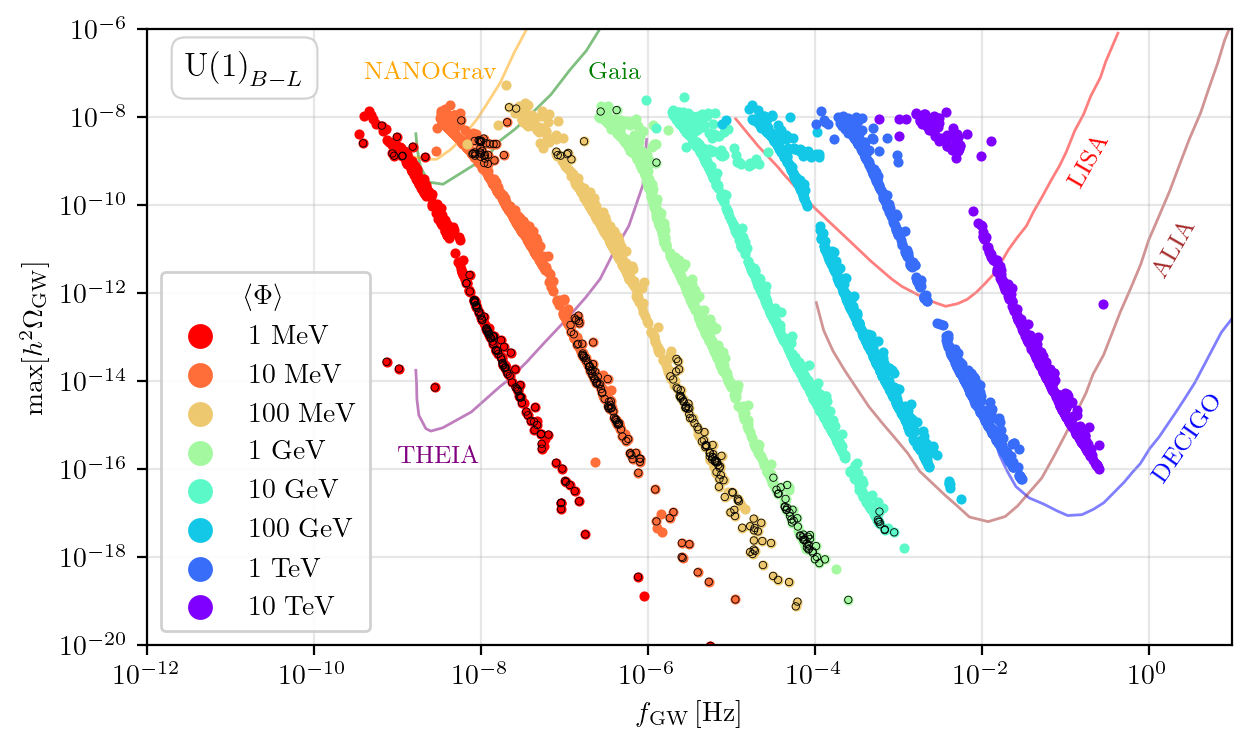}
    \caption{Peaks in GW spectra of the classically conformal U(1)$_{B-L}$ FOPTs found in Fig.~\ref{fig:B-L_FOPT_scan}. Points with black outlines have $f_{\rm PBH} \leq 1$. Colored lines show the projected sensitivities of the following proposed experiments:
     NANOGrav~\cite{NANOGrav:2020bcs, Lazio:2017fos}, Gaia~\cite{brown2018gaia}, THEIA~\cite{2018FrASS...5...11V}, LISA~\cite{Caprini:2019egz, amaro2017laser, Robson:2018ifk}, ALIA~\cite{Gong:2014mca}, DECIGO~\cite{Kawamura:2020pcg, Kawamura:2006up}.} 
    \label{fig:B-L_GW_peaks_VEVs}
\end{figure}

We include the peaks of GW amplitudes for a range of U(1)$_{B-L}$ symmetry-breaking VEVs in Fig.~\ref{fig:B-L_GW_peaks_VEVs}, along with projected sensitivity curves of proposed GW experiments. As the latent heat parameter $\alpha$ is an adimensional parameter independent of $\vevMth$ for conformal potentials, each set of peaks saturates at nearly the same limit of $\max[h^2 \Omega_\mathrm{GW}]$. However, the characteristic timescales of FOPTs, provided by $R_{\rm sep}$, scale with $\vevMth$ and lower VEVs exhibit slower FOPTs, resulting in lower ranges of GW signal frequencies. The main driver of the discrepancies in GW variances between VEVs stems from the temperature sensitivity of the relativistic degrees of freedom $g_R$, which include contributions from the SM and its U(1)$_{B-L}$ extension. Points outlined in black have the additional property of $f_{\rm PBH} \le 1$, and are discussed in more detail in \cref{subsec_PBH_results}.

\begin{figure}[h!]
    \centering
    \includegraphics[width=0.8\textwidth]{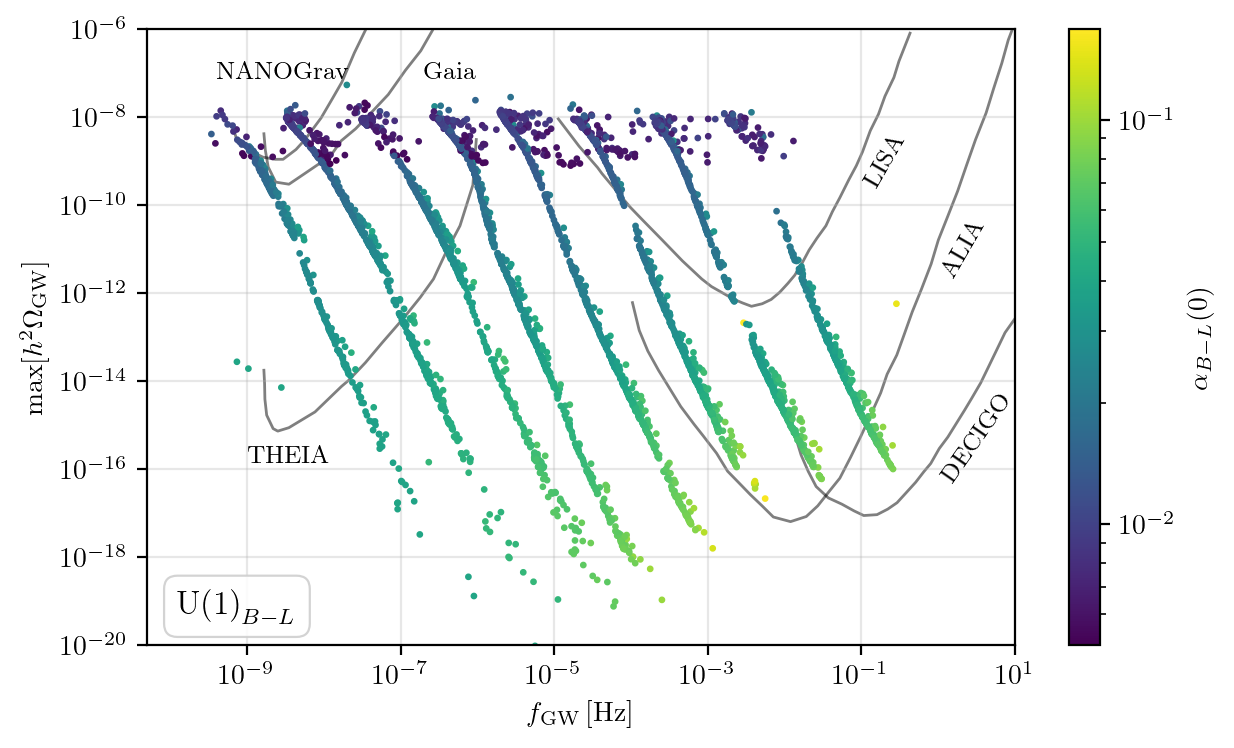}
    \caption{The same data as Fig.~\ref{fig:B-L_GW_peaks_VEVs} with points colored to their corresponding gauge coupling (evaluated at the VEV).}
    \label{fig:B-L_GWs_couplings}
\end{figure}

The same GW peaks, colored according to the gauge coupling evaluated at the VEV, are given in Fig.~\ref{fig:B-L_GWs_couplings}, which shows that models with lower $\alpha_{B-L}(0)$ yield stronger GW signals. To see why, note that GW amplitudes are proportional to $R_{\rm sep}$ (Eqs.~\ref{eq:GW_sound_wave},~\ref{eq:GW_bubble_wall_collision},~\ref{eq:GW_turbulence}). As seen in Fig.~\ref{fig:B-L_FOPT_scan}, lower gauge couplings tend to delay barrier formation in the effective potential, delaying the onset of FOPTs. During this delay, $R_{\rm sep}$ increases from universal expansion, resulting in higher GW amplitudes and stronger signals which are easier to detect. This is in contrast with traditional collider phenomenology, where lower gauge couplings typically yield smaller cross sections of the interactions being studied, making them more difficult to detect and distinguish from background signals. In this way, GW observations provide a complementary probe of particle physics, often gaining sensitivity in regions of parameter space with smaller gauge couplings and potentially higher energy scales that are difficult to access with collider experiments. The combination of GW observations and accelerator searches can therefore yield substantially broader coverage of the underlying parameter space than either approach alone.


\subsection{Predictions: Primordial Black Hole Abundance}~\label{subsec_PBH_results}

Constraints on PBH abundance are typically determined by a lack of observations of expected Hawking radiation or gravitational lensing events~\cite{Carr:2020gox}. Bounds on Hawking radiation help to constrain PBHs with lower mass, as these are expected to emit higher fluxes of higher-energy particles, thus producing stronger signals. Of course, this method struggles to constrain PBHs with larger mass due to their fainter Hawking radiation. Instead, these are better constrained by gravitational lensing searches, as the deflection angles of background light sources are proportional to the mass of the lensing black hole.

The predictions of PBH abundance in the two models considered in this study share some characteristics. Generally, lower VEVs can accommodate higher $\mpbh$ due to the larger Hubble volumes at later times of PBH formation. However, the larger amount of freedom in the generic polynomial potential opens up a much broader phase space with weaker correlations between $\mpbh$ and $\fpbh$. We now turn to contrasting the two models.

\subsubsection{Polynomial Effective Potential}
\begin{figure}[ht]
    \centering
    \includegraphics[width=0.8\linewidth]{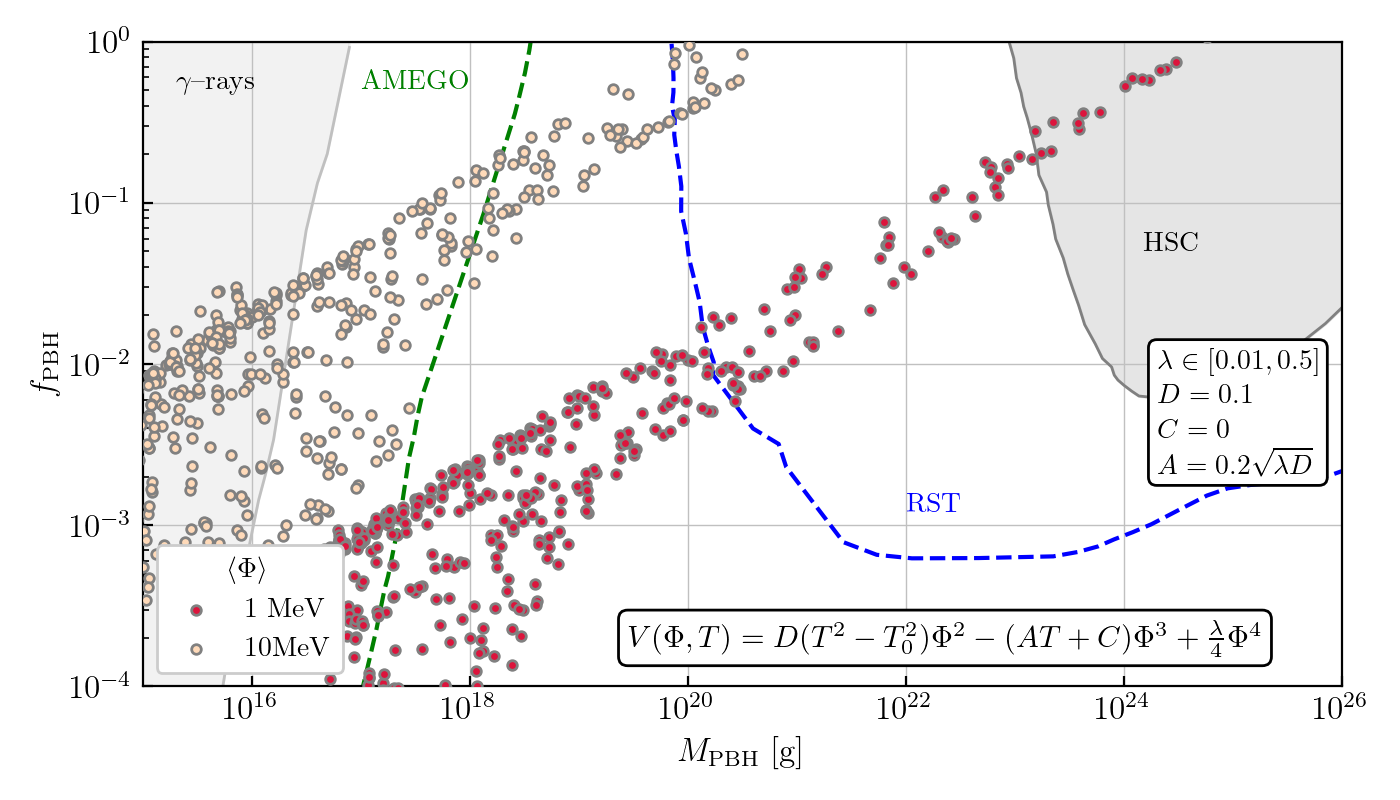}
    \caption{Observational upper bounds on $f_\mathrm{PBH}$ for different PBH mass and $\vevMth$ in the case of the polynomial potential (\cref{eq:generic_pot}).
    Constraints (shown in gray) and future sensitivities (color lines) from the $\gamma$-ray sky test the Hawking evaporation signals in the lower PBH mass limit of the asteroid mass window ($\sim 10^{17} - 10^{23}$~g), while current and future weak lensing searches for non-evaporating PBHs test the upper mass limit.}
    \label{fig:generic_f_PBH_VEVs}
\end{figure}

The PBH abundance predictions for the polynomial potential are shown in Fig.~\ref{fig:generic_f_PBH_VEVs} for fixed VEVs. Due to the larger degree of parameter freedom in the polynomial potential, the spread of PBH abundances can vary broadly, with some models overproducing PBHs and others forming PBHs which are expected to have evaporated before today's epoch through Hawking radiation. Still, a general correlation exists with larger VEVs yielding larger $M_{\rm PBH}$ and $f_{\rm PBH}$, and smaller quartic couplings yielding higher $f_{\rm PBH}$ due to the larger potential well depth induced. Here, we limit the variation in the $(\mpbh, \fpbh)$ plane by restricting the scan to uniform draws over the interval $\lambda \in [0.01, 0.5]$ with $A = 0.2 \times A_{\rm max} = 0.2 \times \sqrt{\lambda D}$ fixed ($D=0.1$). Different choices of the cubic or quadratic couplings will translate into a broader range of $\mpbh$ predictions, owed to the broader access to possible energy densities $\Delta V$ and durations of the phase transitions. 

For the smaller VEV benchmark $\vevMth = 1$~MeV in Fig.~\ref{fig:generic_f_PBH_VEVs}, as $\lambda \to 0.01$ from above, the abundance and PBH mass move upward until the weak lensing bound from Subaru HSC~\cite{Smyth:2019whb} where the abundance reaches nearly 100\% of the DM fraction. Planned lensing searches from RST~\cite{Bird:2022wvk,fardeen2024astrometric,spergel2015widefield} will test a more expansive range down to $\fpbh \sim 10^{-3}$, and for VEVs between $\vevMth \in [1, 10]$~MeV the $\fpbh \sim 1$ PBH DM solution is tested at the tip of the small $\lambda$ limit. Bracketing the lower end of the asteroid mass window ($10^{17}-10^{23}$~g), $\gamma$--ray searches~\cite{Carr:2009jm} and planned observations from AMEGO~\cite{AMEGO:2019gny, Coogan:2020tuf} test for Hawking evaporation signals of the PBH fraction.

\subsubsection{Classically Conformal Effective Potential}
\begin{figure}
    \centering
    \includegraphics[width=0.8\textwidth]{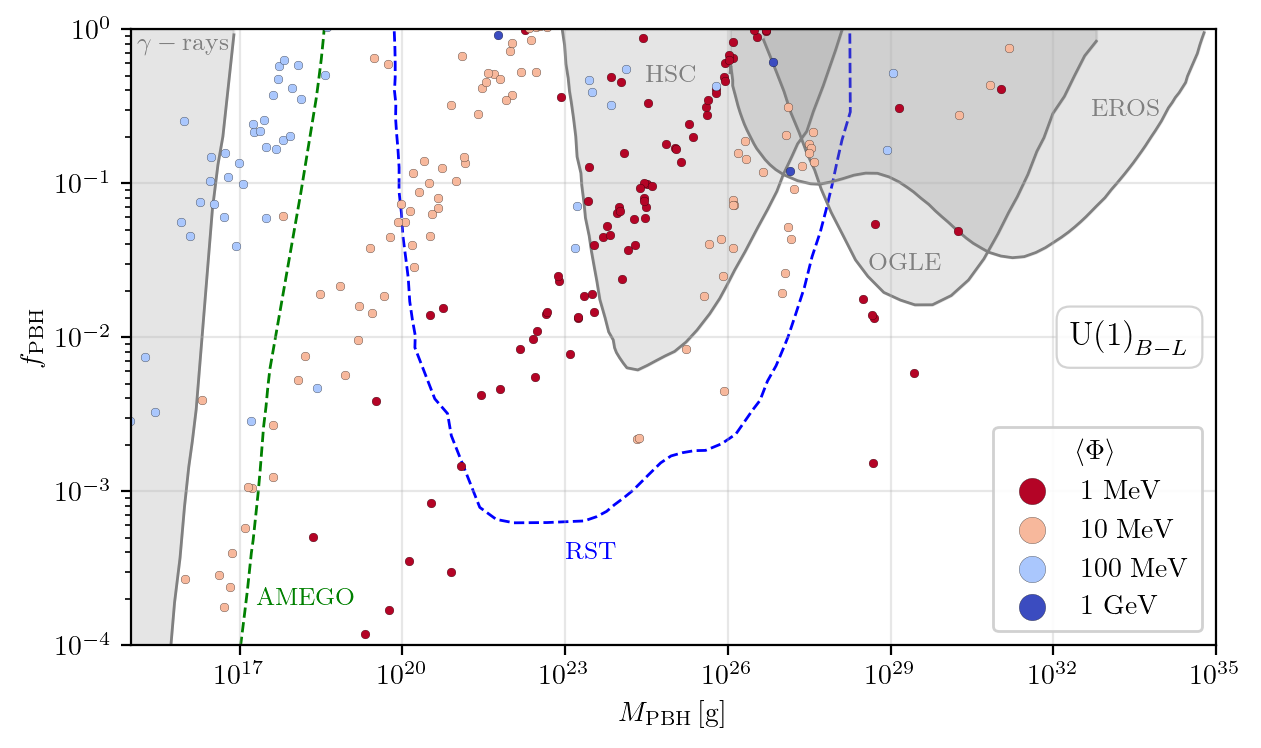}
    \caption{
    Observational upper bounds on $f_\mathrm{PBH}$ for different PBH mass and $\vevMth$ in the case of the classically conformal U(1)$_{B-L}$ effective potential (\cref{eq:B-L_lagrangian}).
    }
    \label{fig:B-L_f_PBH_VEVs}
\end{figure}

Fig.~\ref{fig:B-L_f_PBH_VEVs} offers abundances of PBHs formed in the classically conformal U(1)$_{B-L}$ FOPT.
Shaded regions have been excluded by astrophysical observations, while regions bound from below by colored curves are projected sensitivities of planned experiments. As explained above, models with FOPTs occurring at lower energy scales can accommodate higher $M_{\rm PBH}$ due to additional vacuum energy within larger Hubble volumes. Such models with lower VEVs might then be better constrained by gravitational lensing bounds from HSC in addition to OGLE~\cite{Niikura:2019kqi} or EROS~\cite{EROS-2:2006ryy, Carr:2020gox}, while lensing surveys could supply further probes for these and higher VEVs.
Fig.~\ref{fig:B-L_f_PBH_VEVs} shows exactly this for many points with $\vevMth = 1$~MeV and $\vevMth = 10$~MeV; the latter giving many available models with PBHs making up nearly all the dark matter. Probing U(1)$_{B-L}$ models with higher VEVs (which accommodate lower $M_{\rm PBH}$) is better done with Hawking radiation bounds from $\gamma$--ray searches, with future observations from AMEGO.
Our analysis suggests VEVs near $\vevMth = 100$~MeV can yield models with $f_{\rm PBH} \approx 1$ within AMEGO's expected reach.

Within the parameter space we scan, we also find that large swaths of model parameter space, for both the classically conformal and the polynomial effective potentials, with $\vevMth \gtrsim 1$~GeV either overproduce PBHs ($f_{\rm PBH} > 1$) or form PBHs which are expected to have evaporated before today's epoch through Hawking radiation ($M_{\rm PBH} \lesssim 10^{15}$~g). In principle, several considerations could alter the efficiency of PBH production and tame the abundances; namely, non-sphericity of the false vacuum bubbles would violate the Schwarzschild condition, requiring a more sophisticated collapse criterion that may lower the collapse probability. Secondly, our assumption that the PBH mass function is monochromatic should be weakly broken in principle, since the distribution of false vacuum patch sizes is not uniform albeit highly peaked~\cite{Lu:2022paj}. Going beyond our simplifying assumption that dark radiation does not back-react on the false vacuum bubbles could also play a role in the collapse.

\begin{figure}
    \centering
    \includegraphics[width=0.8\textwidth]{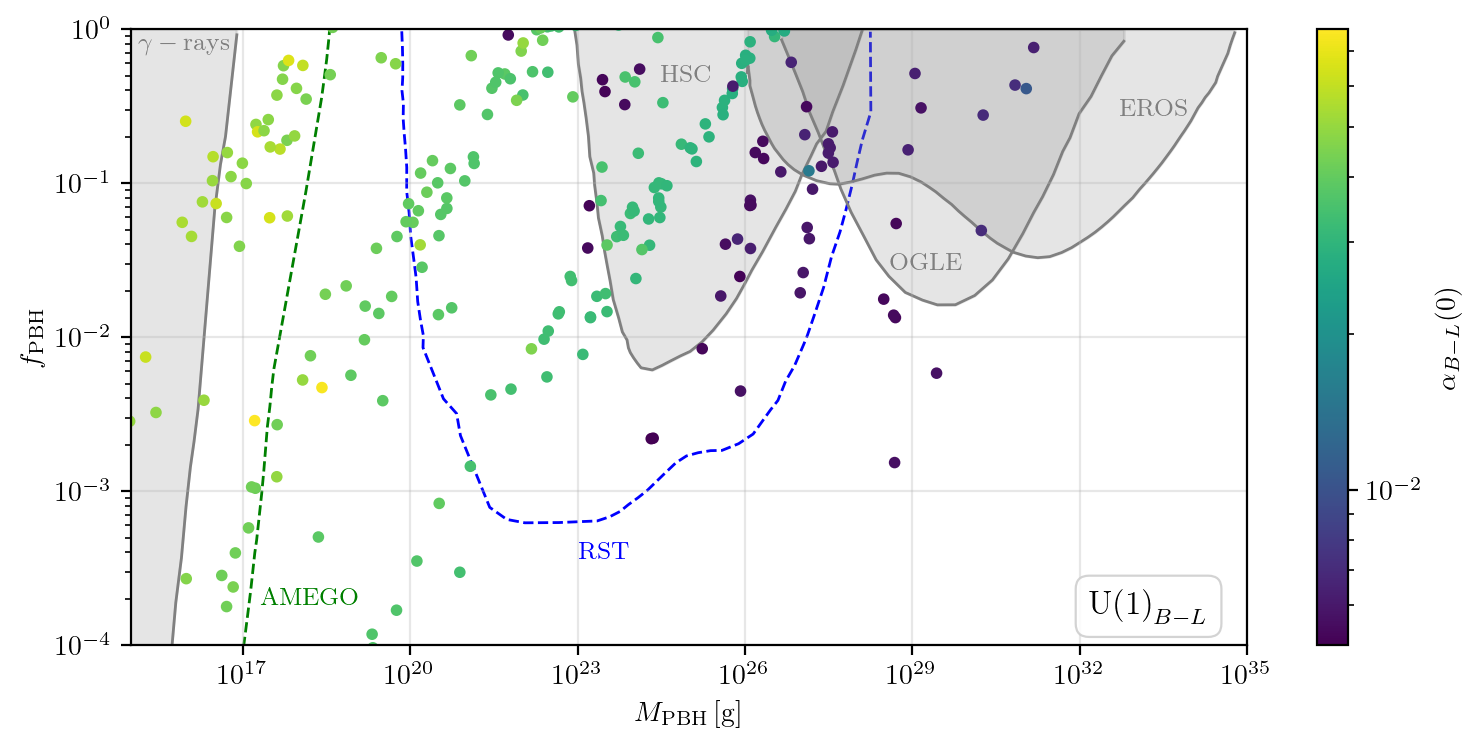}
    \caption{The same data as Fig.~\ref{fig:B-L_f_PBH_VEVs} with points colored to their corresponding gauge coupling (evaluated at the VEV).}
    \label{fig:B-L_f_PBH_couplings}
\end{figure}

Provided these assumptions are taken carefully into consideration, bounds on model parameters can be derived by requiring that the (non-evaporating) PBH abundance today does not exceed the dark matter fraction or lead to excess Hawking radiation. As shown in Fig.~\ref{fig:B-L_f_PBH_couplings}, where we examine the correlation between the U(1)$_{B-L}$ gauge coupling and PBH abundance, we find a trend of increasing $M_{\rm PBH}$ and $f_{\rm PBH}$ as the gauge coupling is decreased. The reasons for this have already been noted: lower gauge couplings tend to delay FOPTs, giving Hubble volumes time to expand and collapse to PBHs of higher mass. A major discontinuity in this trend is found in models with small gauge couplings of $\alpha_{B-L}(0) \lesssim 10^{-2}$; comparing Figs.~\ref{fig:B-L_f_PBH_VEVs} and \ref{fig:B-L_f_PBH_couplings} reveals that these points tend to leave the main regions inhabited by their VEVs. As expected, the lower gauge couplings yield higher $M_{\rm PBH}$, but these models are able to avoid PBH overproduction due to their significant reheating (characterized by large values of $\alpha$). As seen in Eq.~\eqref{eq:f_PBH_fraction_today}, such an injection of entropy dilutes the number density of PBHs and therefore diminishes $f_{\rm PBH}$. This behavior of both large and small $\alpha_{B-L}(0)$ values yielding $f_{\rm PBH} \leq 1$ is also apparent in Figs.~\ref{fig:B-L_GW_peaks_VEVs} and \ref{fig:B-L_GWs_couplings}.

\subsection{Multi-messenger Probes}
\label{sec:multimessenger}
\subsubsection{Polynomial Effective Potential}
\begin{figure}[h]
    \centering
    \includegraphics[width=1.0\textwidth]{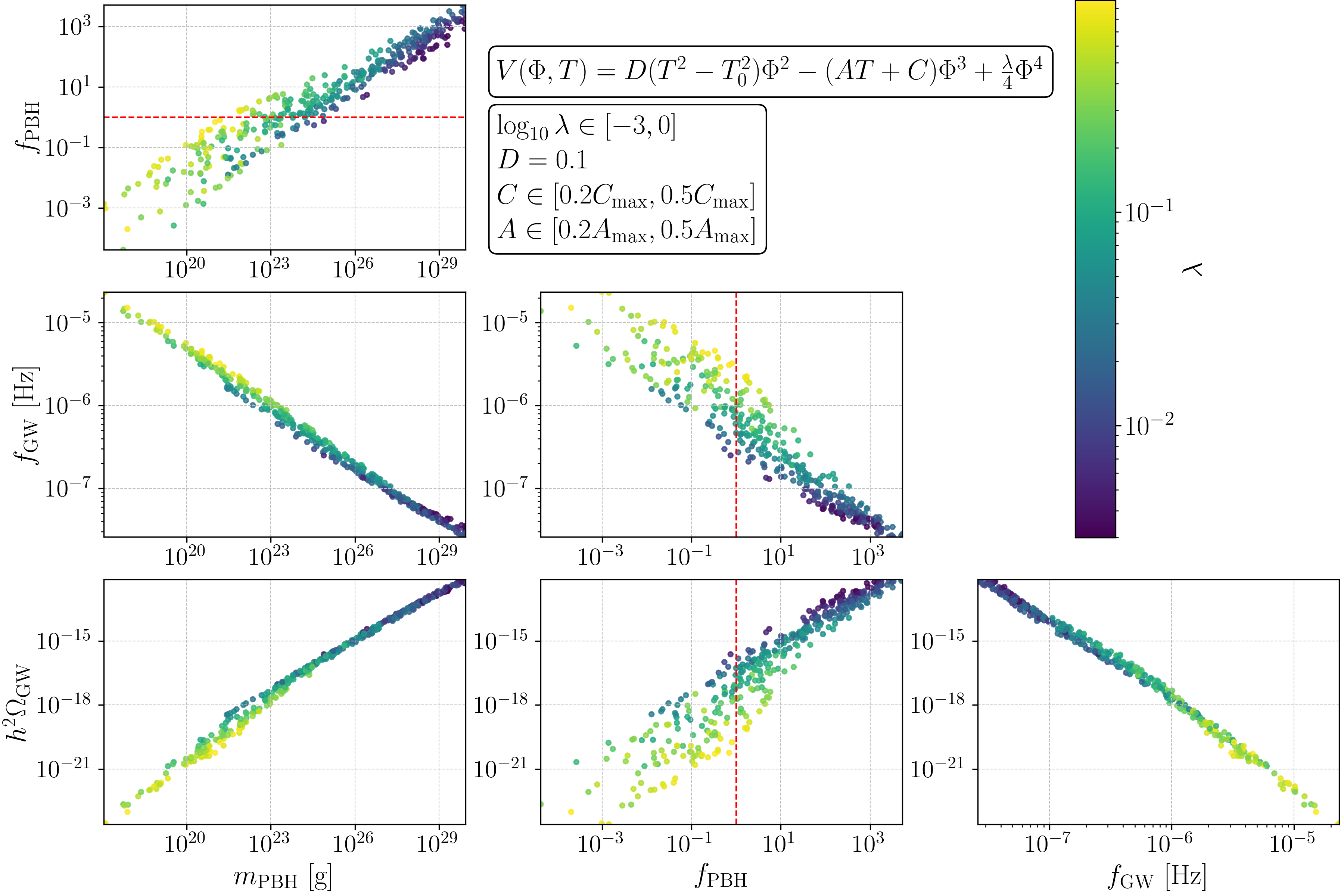}
    \caption{Parameter scans for the polynomial potential with $\vevMth = 1$ MeV. We show the distribution of points in the space of GW and PBH observables for points in model parameter space (with $D=0.1$ and uniform distributions drawn for $\log_{10}\lambda \in [-3, 0]$, $C \in [0.2 C_{\rm max}, 0.5 C_{\rm max}]$, and $A \in [0.2 A_{\rm max}, 0.5 A_{\rm max}]$, where $C_{\rm max} = v \lambda / 3$ and $A_{\rm max} = \sqrt{\lambda D}$) that gave rise to first-order phase transitions, color coded by the size of the quartic coupling $\lambda$. Correlations across all four observables are discussed in the text. The dashed red line indicates where $\fpbh > 1$.}
    \label{fig:parameter_scans_1MeV}
\end{figure}

We now turn to investigating correlated observables in the production of PBH and GW signals from the first order phase transitions we considered in the previous sections. Beginning first with the polynomial potential for fixed \vev$=1$ MeV, in Fig.~\ref{fig:parameter_scans_1MeV} we examine correlations between the GW frequency $f_{\rm GW}$ and peak strain $\max [h^2 \Omega_{\rm GW}]$ against the PBH masses and abundances formed by vacuum collapse (if PBH formation fails, we assign an abundance of zero). We color code the points in Fig.~\ref{fig:parameter_scans_1MeV} by the value of the quartic coupling $\lambda$ in the random scan. 

We can explain the correlations between the four GW and PBH observables in each panel of Fig.~\ref{fig:parameter_scans_1MeV} as follows. Setting aside temperature dependence, which is roughly fixed by the scale of the vev $\vevMth$, since $\fgw$ grows with the inverse power of $\rsep$ and the peak gravitational strain of $\hto$ grows with $\rsep$ (as in \cref{eq:GW_sound_wave}), we know that $\fgw$ and $\hto$ are inversely proportional through their power law dependence on $\rsep$. On the other hand, $\mpbh \propto \rsep^3$ via \cref{eq:PBH_mass_Misner_Sharp} since the volume term dominates the contribution to the PBH mass. The abundance $\fpbh \propto \mpbh n_{\rm PBH}$ (\cref{eq:rho_PBH_formation_PT}), and the size of $n_{\rm PBH}$ is largely set by the size of the rate $\Gamma_f$ and $\Gamma(T)$ (\cref{eq:lu_rate} and \cref{eq:tunneling_probability_rate}). Since these rates depend most sensitively on the temperature with positive correlation, for a fixed temperature scale we have $\fpbh \propto \mpbh \propto \rsep^3$ via the same growth with $\rsep$. Then, we estimate very roughly,
\begin{equation}
    \fpbh \propto \mpbh \propto \big(\fgw \big)^{-1/3} \propto \big(\hto \big)^{1/3} \, .
\end{equation}
This reflects the correlations seen in \cref{fig:parameter_scans_1MeV}.


\subsubsection{Classically Conformal Effective Potential}
\begin{figure}
    \centering
    \includegraphics[width=0.8\linewidth]{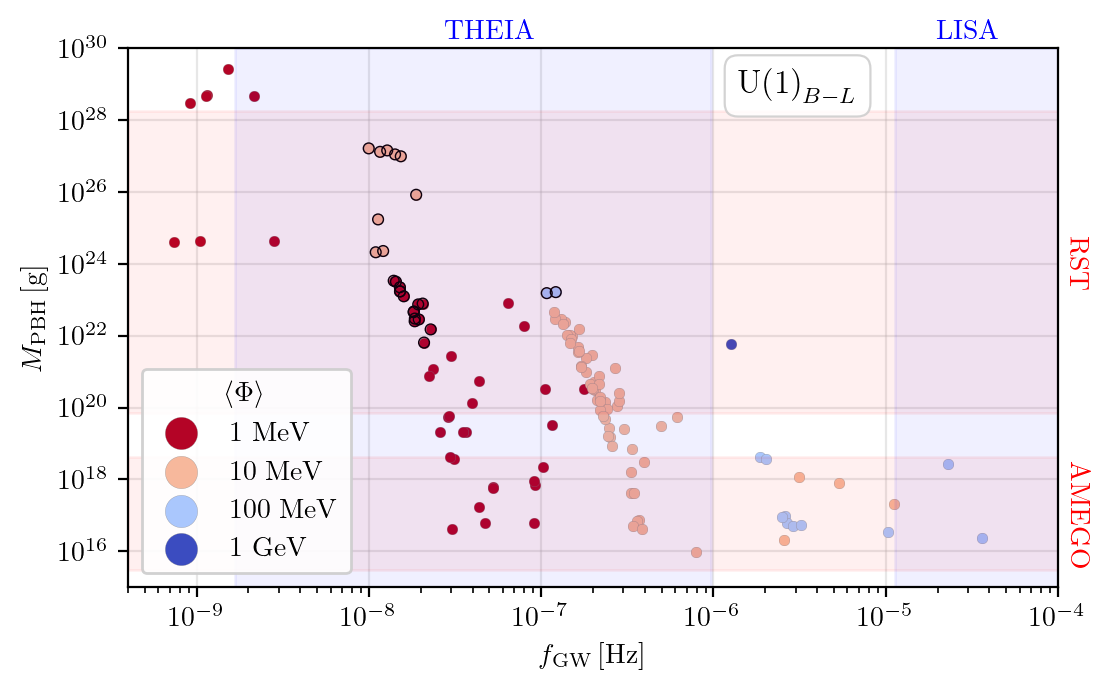}
    \caption{Multi-messenger parameter space of scanned U(1)$_{B-L}$ models exhibiting FOPTs favorable to PBH formation. Blue shaded regions give the expected frequency ranges to which THEIA and LISA are sensitive. Red shaded regions give the expected ranges of PBH masses to be probed by RST and AMEGO. Points with black outlines may be simultaneously probed by proposed GW detectors and PBH searches.}
    \label{fig:B-L_fpeak_vs_M}
\end{figure}

We now offer a look at our collection of classically conformal U(1)$_{B-L}$ models in a multi-messenger parameter space. Figs.~\ref{fig:B-L_GW_peaks_VEVs} and \ref{fig:B-L_f_PBH_VEVs} make it clear that models with different VEVs $\vevMth$ fall under the purview of different experiments. The same data are presented in Fig.~\ref{fig:B-L_fpeak_vs_M} to show which models could be simultaneously probed by separate experiments of GW detection and $f_\mathrm{PBH}$ observation. In these figures, the scanned FOPT points favorable to PBH formation are plotted according to the frequencies $f_\mathrm{GW}$ at which their GW signal is maximized and the PBH mass $M_\mathrm{PBH}$. There are four shaded regions which demarcate sensitivities of proposed experiments. The two blue regions cover the GW frequency ranges expected to be investigated by THEIA and LISA, while the two red regions carve out PBH masses within reach of the proposed RST and AMEGO experiments. Of course, in determining whether a given U(1)$_{B-L}$ model is testable by any of these experiments, the model's GW signal peak $\max[h^2 \Omega_\mathrm{GW}]$ and PBH fraction $f_\mathrm{PBH}$ must be considered (as in Figs.~\ref{fig:B-L_GW_peaks_VEVs} and \ref{fig:B-L_f_PBH_VEVs}). Those special models which may be simultaneously probed by at least one proposed GW detector and at least one PBH search are outlined in black.


The combination of higher PBH masses and lower $f_\mathrm{PBH}$ that is expected from the classically conformal models with lower energy scales suggests these models provide the best candidates to be explored by both GW detection and $f_\mathrm{PBH}$ observations (without already being excluded by the latter). In particular, this analysis finds many models with VEVs from 1 MeV to 100 MeV whose PBH fraction can be probed by future microlensing surveys done by the RST, and whose GW spectra can be simultaneously probed by THEIA. We provide three such models (BM1, BM2, and BM3) as benchmarks in Table~\ref{tab:multi-messenger_B-L_benchmark_points}. Benchmarks BM4, BM5, BM6, and BM7 are models whose formed PBHs are abundant enough to make up most or all of our universe's dark matter. Note the significant reheating in BM7 (marked by a high $\alpha$); as mentioned in \cref{subsec_PBH_results}, dilution in $f_\mathrm{PBH}$ from reheating is often necessary for models with higher VEVs to avoid PBH overproduction. Finally, the models given by BM8 and BM9 fit into both of the benchmark categories outlined above: they (1) may be reached simultaneously by proposed GW detectors and PBH searches and (2) provide all observed dark matter in the form of PBHs formed near the end of their FOPT.

\begin{table}[h!]
    \centering
    \caption{Classically conformal U(1)$_{B-L}$ benchmark models. The first three (BM1, BM2, BM3) are simultaneously probeable by proposed GW detectors and PBH searches; specifically THEIA and RST. The next four (BM4, BM5, BM6, BM7) allow for the majority of dark matter to be PBHs. The last two points (BM8, BM9) may be simultaneously probed by RST and THEIA as well as provide all dark matter in the form of PBHs.}
    \begin{tabular}{c c c c|c c c|c c c}
        \hline \hline
         & $\vevMth$ [MeV] & $\alpha_{B-L}(0)$ & $\alpha_{Y}(0)$ & $T_\mathrm{perc}$ [MeV] & $\alpha$ & $R_{\rm sep} H(T_\mathrm{perc})$ & $f_\mathrm{GW}$ [Hz] & $M_{\rm PBH}$ [g] & $f_{\rm PBH}$ \\
        \hline
        BM1 & 1 & 0.039 & 0.095 & 0.37 & 0.13 & 0.0039 & $1.9 \times 10^{-8}$ & $3.4 \times 10^{23}$ & 0.040 \\

        BM2 & 10 & 0.0061 & 0.0024 & 0.038 & $7.6 \times 10^{6}$ & 0.039 & $9.9 \times 10^{-9}$ & $3.8 \times 10^{27}$ & 0.14 \\

        BM3 & 100 & 0.0054 & 0.0042 & 0.23 & $2.0 \times 10^{7}$ & 0.028 & $1.1 \times 10^{-7}$ & $1.5 \times 10^{23}$ & 0.038 \\
        \hline
        BM4 & 1 & 0.038 & $1.0 \times 10^{-5}$ & 0.50 & 0.033 & 0.0012 & $8.0 \times 10^{-8}$ & $1.9 \times 10^{22}$ & 0.99 \\

        BM5 & 10 & 0.042 & 0.13 & 3.4 & 0.18 & 0.0052 & $1.3 \times 10^{-7}$ & $2.2 \times 10^{22}$ & 1.0 \\

        BM6 & 100 & 0.043 & $3.0 \times 10^{-4}$ & 43 & 0.065 & 0.0047 & $1.9 \times 10^{-6}$ & $4.2 \times 10^{18}$ & 1.0 \\

        BM7 & $10^3$ & 0.0054 & $9.4 \times 10^{-4}$ & 2.3 & $1.6 \times 10^7$ & 0.025 & $1.3 \times 10^{-6}$ & $5.7 \times 10^{21}$ & 0.92 \\
        \hline
        BM8 & 5 & 0.032 & $9.5 \times 10^{-4}$ & 1.7 & 0.14 & 0.0059 & $5.7 \times 10^{-8}$ & $3.5 \times 10^{23}$ & 1.0 \\

        BM9 & 8 & 0.041 & 0.13 & 2.7 & 0.20 & 0.0055 & $9.9 \times 10^{-8}$ & $8.0 \times 10^{22}$ & 1.0 \\
        \hline \hline
    \end{tabular}
    \label{tab:multi-messenger_B-L_benchmark_points}
\end{table}

\section{Conclusions}
\label{sec:conclusion}

We have studied cosmological first-order phase transitions and their multi-messenger signatures in two complementary settings: a generic quartic polynomial potential with four parameters, and a classically conformal U(1)$_{B-L}$-breaking potential whose dynamics are controlled by two running couplings and thermal corrections. Using a modified version of \texttt{ELENA} to compute the phase transition characteristics, we mapped the resulting stochastic GW signals and, through the Israel junction-condition description of false-vacuum collapse, the corresponding PBH masses and abundances across the scanned parameter space.
In contrast to conventional PBH production through overdensity collapse, the false-vacuum collapse mechanism does not rely on particle trapping or overdensity thresholds. Instead, PBH formation follows directly from the Schwarzschild criterion applied to the false-vacuum domains at the end of a FOPT~\cite{Flores:2024lng}. This treatment provides a directly calculable connection between the particle physics governing the phase transition and the resulting PBH population.

Our main phenomenological result is the identification of a viable multi-messenger region in parameter space, as shown in Fig.~\ref{fig:B-L_fpeak_vs_M} and instantiated by the benchmarks of Table~\ref{tab:multi-messenger_B-L_benchmark_points}. For classically conformal U(1)$_{B-L}$ extensions of the SM, the most promising region for combined GW and PBH searches sits within
$\vevMth \in [1, 10^3]$~MeV
, where GW peak frequencies fall within the projected sensitivity windows of THEIA and LISA, while the corresponding PBH masses
$M_{\rm PBH} \in [10^{16}, 10^{28}]$~g
lie within the expected reach of RST microlensing, and in some cases of AMEGO $\gamma$--ray searches. Higher VEVs ($\vevMth \gtrsim 1$~GeV) are less favorable for multi-messenger searches because the efficient PBH formation at these energy scales tends to yield populations which either overproduce dark matter ($f_{\rm PBH} > 1$) or generate masses below the evaporation cutoff $M_{\rm PBH} \approx 10^{15}$~g. This overproduction appears in both the polynomial and classically conformal effective potentials, although it may be mitigated by more realistic false-vaccum radius distributions or non-sphericity, or by the presence of dark radiation that may drive pressure instabilities during trapping~\cite{Lewicki:2023mik}.

We identify a preferred region of parameter space in classically conformal models with $1~\text{MeV} \lesssim \langle \Phi \rangle \lesssim 10~\text{MeV}$, where the phase transition simultaneously produces observable GWs and asteroid-mass PBHs with $f_\mathrm{PBH} \sim 1$, including models in which PBHs constitute essentially all of the dark matter. Below this range, the GW spectrum is shifted to frequencies below about $10^{-9}$~Hz (beyond the reach of currently proposed GW experiments), while the corresponding PBHs become sufficiently massive to be constrained by gravitational lensing surveys (with additional considerations arising from big bang nucleosynthesis at and below these symmetry-breaking scales~\cite{Cyburt:2015mya,Sabti:2019mhn,Pospelov:2010hj}). Above this range, progressively lighter PBHs are instead constrained by searches for Hawking radiation---if not already evaporated before the present epoch. Consequently, classically conformal symmetry breaking at MeV-scales seem to provide the most promising targets (such as BM8 and BM9 in Table~\ref{tab:multi-messenger_B-L_benchmark_points}) for multi-messenger probes of PBH formation through vacuum collapse.


We also emphasize the complementarity of GW observations with traditional collider searches for new physics. Whereas colliders probe interactions up to $\mathcal O(\text{TeV})$ energies, the GW signals of classically conformal FOPTs retain their characteristic amplitudes over many orders of magnitude in \vev because $\alpha$ is approximately scale-invariant for such potentials. This opens a window into symmetry-breaking scales that are complementary to the reach of accelerators,  colliders, and low-energy probes~\cite{Bandyopadhyay:2018cwu, Wang:2024gvt, DeRomeri:2024iaw, Deppisch2019SearchingFA}. At much higher scales, phenomena such as string formation can still give rise to cosmological observables, but are well out of reach of accelerator complementarity~\cite{Ghoshal:2025iil}; within the potential scans, the strongest GW signals actually arise at the lowest gauge couplings (in the classically conformal case) or lower quartic couplings (in the polynomial effective potential) which effectively lowers the cubic barrier and deepens the potential well. GW observations can therefore probe parts of parameter space that are intrinsically invisible to high-energy collider programs via this sensitivity to small couplings that affect the potential shape. 

We find that model dependence in the phase transition potential is reflected in the resulting distributions in GW and PBH phase space: the classically conformal potential and the polynomial potential benchmarks both trace out trajectories in the $(\mpbh, \fpbh, \fgw, \hto)$ parameter space. The range of these observables can grow dramatically with added model freedom, which we find in the case of the polynomial potential that enjoys a parametric freedom via independent quadratic, cubic, and quartic couplings. For more physical models, like the U(1)$_{B-L}$ realization of a classically conformal potential we studied here, the potential parameters are more strongly correlated and determined by fewer parameters. This U(1) model benchmark can be connected to accelerator and other laboratory probes in a model-dependent way; other choices of U(1) may be more or less constrained in the multi-messenger regions of parameter space.

Searches for PBHs formed in FOPTs can further extend the broader program of using cosmological and astrophysical observables to search for new particle physics. In the PBH formation mechanism considered here (originated in ref.~\cite{Blau:1986cw}), the connection between the underlying particle physics and the resulting PBH population is relatively transparent, as PBH formation is determined by a turning-point condition for vacuum bubbles whose properties are fixed by the phase transition. Although this treatment neglects potentially important interactions between bubble walls and the surrounding plasma, the models identified in our scans predominantly reside in a vacuum-dominated regime at the time of PBH formation, suggesting that such corrections may be subleading. Moreover, we found that PBH searches are not merely redundant with GW observations. While the strongest GW signals in our classically conformal models generally arise from smaller gauge couplings---largely independent of \vev, PBH abundance can exhibit a different dependence on model parameters. Specifically, Fig.~\ref{fig:B-L_f_PBH_couplings} shows how lower symmetry-breaking scales can accommodate somewhat larger gauge couplings while still producing larger $f_\mathrm{PBH}$. Hence, regions of parameter space yielding weaker GW signals may nevertheless generate observable PBH populations.


\section*{Acknowledegments}
We thank Mudit Rai, Tao Xu, Joachim Kopp, and Isaac Wang for the fruitful discussions and feedback. We also thank Martin Vasar for the insightful technical feedback in supplemental explorations using the code \textsc{bubbleSim}~\cite{Lewicki:2023mik}. The work of AT is supported in part by the U.S. DOE grant \#DE-SC0010143.
The work of BD is supported in part by the U.S. DOE grant \# DESC0010813. The work of CH and PH is supported by the National Science
Foundation under grant number PHY-2412875. The work of PH is also supported by the University Research Association Visiting Scholars Program. PH is grateful for the hospitality of Perimeter Institute where part of this work was carried out. Research at Perimeter Institute is supported in part by the Government of Canada through the Department of Innovation, Science and Economic Development and by the Province of Ontario through the Ministry of Colleges and Universities. This work was supported by a grant from the Simons Foundation (1034867, Dittrich). We are grateful to the Center for Theoretical Underground Physics and Related Areas (CETUP*), The Institute for Underground Science at Sanford Underground Research Facility (SURF), and the South Dakota Science and Technology Authority for their hospitality and financial support.


\appendix

\section{Gravitational Wave Spectra: Fit Functions}
\label{app:gwfits}
Parameterizations of the GW spectra from FOPTs can be described with the following fits for contributions from sound waves in the plasma (sw), collisions of vacuum-bubble walls (col), and magnetohydrodynamic turbulence (turb) within the plasma:
\begin{equation}
    \label{eq:GW_sound_wave}
    \begin{split}
        h^2 \Omega_{\rm sw} &= 0.38 [H(T_p) R_{\rm sep}] [H(T_p) \tau_{\rm sw}] \left( \frac{\kappa_{\rm sw} \alpha}{1 + \alpha} \right)^2 \left( \frac{f}{f_{\rm sw}} \right)^3 \left[ 1 + \frac{3}{4} \left( \frac{f}{f_{\rm sw}} \right)^2 \right]^{-\frac{7}{2}} \times r_\Omega \, ,\\
        f_{\rm sw} &= \frac{1.\overline{57}}{R_{\rm sep}} \times r_f \, ,
    \end{split}
\end{equation}
\begin{equation}
    \label{eq:GW_bubble_wall_collision}
    \begin{split}
        h^2 \Omega_{\rm col} &= 0.024 [H(T_p) R_{\rm sep}]^2 \left( \frac{\kappa_{\rm col} \alpha}{1 + \alpha} \right)^2 \left( \frac{f}{f_{\rm col}} \right)^3 \left[ 1 + 2 \left( \frac{f}{f_{\rm col}} \right)^{2.07} \right]^{-2.18} \times r_\Omega \, , \\
        f_{\rm col} &= \frac{0.51}{R_{\rm sep}} \times r_f \, ,
    \end{split}
\end{equation}
\begin{equation}
    \label{eq:GW_turbulence}
    \begin{split}
        h^2 \Omega_{\rm turb} &= 6.8 [H(T_p) R_{\rm sep}] [1 - H(T_p) \tau_{\rm sw}] \left( \frac{\kappa_{\rm sw} \alpha}{1 + \alpha} \right)^{3/2} \frac{\left( \frac{f}{f_{\rm turb}} \right)^3 \left[ 1 + \left( \frac{f}{f_{\rm turb}} \right) \right]^{-11/3}}{1 + 8 \pi f / H(T_p)} \times r_\Omega \, , \\
        f_{\rm turb} &= \frac{3.9}{(v_w - c_s) R_{\rm sep}} \times r_f \, .
    \end{split}
\end{equation}
The factors $r_\Omega$ and $r_f$ are used to redshift the GW signal's amplitude and frequency to today's values:
\begin{equation}
    \label{eq:GW_signal_redshift}
    \begin{split}
        r_\Omega &:= 1.67 \times 10^{-5} \left(\frac{100}{g_R(T_{\rm reh})}\right)^{1/3} \, , \\
        r_f &:= 1.65 \times 10^{-5} \, \mathrm{Hz} \, \left(\frac{T_{\rm reh}}{100 \, \mathrm{GeV}}\right) \left(\frac{100}{g_R(T_{\rm reh})}\right)^{-1/6} \frac{1}{H(T_p)} \, ,
    \end{split}
\end{equation}
with $h$ the reduced Hubble parameter today: $H_0 = 100 \, h \, \mathrm{km} \, \mathrm{s}^{-1} \, \mathrm{Mpc}^{-1}$.
To determine each of these contributions, one needs knowledge of the fractional amount of energy released by the FOPT that transfers into the bubble wall kinetic energy ($\kappa_\mathrm{col}$) and the bulk motion of the fluid ($1 - \kappa_\mathrm{col}$). These ``efficiency parameters'' are found to be~\cite{Ellis:2019oqb}
\begin{equation}
    \label{eq:efficiency_parameters}
    \begin{split}
    \kappa_{\rm col} &=
    \begin{cases} 
   \frac{\gamma_{\rm eq}}{\gamma_*} \left[ 1 - \frac{\alpha_\infty}{\alpha} \left( \frac{\gamma_{\rm eq}}{\gamma_*} \right)^2 \right], & \gamma_* > \gamma_{\rm eq} \\
   1 - \frac{\alpha_\infty}{\alpha}, & \gamma_* \leq \gamma_{\rm eq}
    \end{cases} \\
    \kappa_{\rm sw} &= \frac{\alpha_{\rm eff}}{\alpha} \frac{\alpha_{\rm eff}}{0.73 + 0.083\sqrt{\alpha_{\rm eff}} + \alpha_{\rm eff}} , \quad \alpha_{\rm eff} := \alpha(1 - \kappa_{\rm col}).
    \end{split}
\end{equation}
The parameter $\alpha_\infty$ is the minimal $\alpha$ required to overcome leading-order friction effects between the vacuum domain walls and particles whose mass changes across the walls~\cite{Espinosa:2010hh}:
\begin{equation}
        \alpha_\infty = \frac{T_p^2}{18 w(\phi_\mathrm{false}, T_p)} \sum_i c_i n_i \left[m_i(\phi_\mathrm{true})^2 - m_i(\phi_\mathrm{false})^2\right] \, .
\end{equation}
Here, $c_i$ is 1 for bosons and 1/2 for fermions, and $n_i$ counts their internal degrees of freedom.
When domain walls become relativistic, next-to-leading-order friction becomes significant and is quantified with $\alpha_\mathrm{eq}$~\cite{Bodeker:2009qy,Bodeker:2017cim,Gouttenoire:2021kjv},
\begin{equation}
        \alpha_\mathrm{eq} = \frac{4 T_p^3}{3 w(\phi_\mathrm{false}, T_p)} \sum_V 3 g_V^2 \left[m_V(\phi_\mathrm{true}) - m_V(\phi_\mathrm{false})\right] \, ,
\end{equation}
which sums over vector gauge bosons and their gauge couplings $g_V$.
The Lorentz factor,
\begin{equation}
    \gamma_\mathrm{eq} = \frac{\alpha - \alpha_\infty}{\alpha_\mathrm{eq}} \, ,
\end{equation}
describes the terminal speed reached by bubble walls from next-to-leading-order friction~\cite{Bodeker:2017cim,Gouttenoire:2021kjv}. For a bubble wall starting from rest at radius $R_0$ and expanding until it collides with radius $R_*$, the Lorentz factor is
\begin{equation}
    \gamma_* = \frac{2 R_*}{3 R_0} \, ,
\end{equation}
which is reached in the absence of next-to-leading-order friction~\cite{Ellis:2019oqb}.
In the above GW strains, $\tau_{\rm sw}$ gives an estimate to the lifetime of the sound waves as
\begin{equation}
    \tau_{\rm sw} \equiv \min\bigg[\frac{1}{H(T_p)}, \frac{R_{\rm sep}}{U_f} \bigg] \, ,
\end{equation}
where the RMS fluid velocity is approximated as $U_f^2 \simeq \frac{3}{4} \frac{\alpha_{\rm eff}}{1 + \alpha_{\rm eff}} \kappa_{\rm sw}$.

In the models presented in this analysis, sound waves are (by far) the dominant contributors to GW spectra for two main reasons. The first comes from next-to-leading-order friction causing bubble walls to saturate in their speeds ($\gamma_* \gg \gamma_\mathrm{eq}$), leading to an efficient conversion of energy from the moving walls to the surrounding plasma. Combining this with the high values of $\alpha$ leads to $\kappa_\mathrm{col} \ll 1$. The second reason is that $\tau_\mathrm{sw} > R_\mathrm{sep}$ for our scanned models, indicating a relative enhancement of $h^2 \Omega_\mathrm{sw}$ from the prolonged lifetimes of sound waves in the plasma.
Note also that in the absence of fermion mass generation in the true vacuum, $\alpha_\infty \to 0$.
This holds for the generic scalar potential which we considered in this work without Yukawa couplings to fermions. In this case, we omit the term $\kappa_{\rm col}$ accounting for friction on the bubble wall, taking $\alpha_{\rm eff} \to \alpha$.

\section{Validity of Eq.~\eqref{eq:PBH_mass_Misner_Sharp} in Flat FLRW Manifolds}
\label{app:BGG_validity}
Consider a three-dimensional hypersurface $\Sigma$ embedded in a four-dimensional spacetime manifold with metric components $g_{\mu\nu}$. Let $s^\mu$ be a unit normal form to $\Sigma$. The induced metric $h$ and extrinsic curvature $K$ of $\Sigma$ have the components
\begin{equation}
\begin{split}
    h_{\mu\nu} &= g_{\mu\nu} - \epsilon s_\mu s_\nu \, ,\\
    K_{\mu\nu} &= h_\mu^{\;\; \lambda} \nabla_\lambda s_\nu \, ,
\end{split}
\end{equation}
with $\epsilon \equiv s_\mu s^\mu = +1$ when $\Sigma$ is spacelike, and $\nabla$ is the covariant derivative. The Israel junction conditions read~\cite{Blau:1986cw}
\begin{equation}
\begin{split}
    h_{\mu\nu}^+ - h_{\mu\nu}^- &= 0 \, ,
    \\
    K_{\mu\nu}^+ - K_{\mu\nu}^- &= 8\pi G \epsilon \left( -S_{\mu\nu} + \frac{1}{2} S h_{\mu\nu} \right) \, ,
\end{split}
\end{equation}
where $S_{\mu\nu}$ are components of the energy-momentum tensor of fields on $\Sigma$ and $S \equiv h_{\mu\nu} S^{\mu\nu}$. The $\pm$ superscripts denote values evaluated on opposite sides of $\Sigma$. In ref.~\cite{Blau:1986cw} Blau, Guendelman, and Guth (BGG) derive Eq.~\eqref{eq:PBH_mass_Misner_Sharp} through the junction condition on $K_{\theta\theta}$ at the interface of a de Sitter manifold and a Schwarzschild manifold. We will now provide conditions which allow their results to approximately hold valid on an interface between two flat FLRW manifolds with a difference in vacuum energy. Assuming particle interactions with domain walls separating vacuums are negligible, $S_{\mu\nu}$ are not affected by this change in manifolds, and we need only to look at changes in the extrinsic curvature.

For any spherically symmetric manifold, the nonzero metric components may be generally written as
\begin{equation}
    g_{tt} = -e^{2\alpha(t,\chi)} \, ,
    \quad
    g_{\chi\chi} = e^{2\beta(t,\chi)} \, ,
    \quad
    g_{\theta\theta} = R(t,\chi)^2 \, ,
    \quad
    g_{\phi\phi} = \left[R(t,\chi) \sin(\theta) \right]^2 \, .
    \quad
\end{equation}
Within a FOPT, we take the domain walls to be spherical shells moving on $\Sigma$ with coordinates $\left(t^s(\tau), \chi^s(\tau)\right)$ where $\tau$ is the shell's proper time. The shell's four-velocity then has components
\begin{equation}
    v^\mu = \left(\Dot{t}^s, \Dot{\chi}^s, 0, 0\right) \, ,
\end{equation}
which are related through the normalization of four-velocities:
\begin{equation}
\label{eqB:four-velocity_normalization}
    -1 = v_\mu v^\mu
    \implies
    e^{4\alpha} \left(\Dot{t}^s\right)^2 - e^{2(\alpha+\beta)} \left(\Dot{\chi}^s\right)^2 = e^{2\alpha} \, ,
\end{equation}
where dots indicate derivatives with respect to proper time $\tau$. As any object's velocity is tangent to the surface on which it moves, the shell's four-velocity defines a radial tangent vector on $\Sigma$ such that $v_\mu s^\mu = 0$. By combining this condition with the normalization $s_\mu s^\mu = 1$, and appealing to spherical symmetry to set $s^\theta = s^\phi = 0$, one can derive all four components of the unit normal form to find
\begin{equation}
    s_\mu = \pm e^\alpha e^\beta \left(-\Dot{\chi}^s, \Dot{t}^s, 0, 0\right) \, ,
\end{equation}
which has been simplified with Eq.~\eqref{eqB:four-velocity_normalization}.

We can now simplify the components of the extrinsic curvature of a spherically symmetric manifold. In deriving Eq.~\eqref{eq:PBH_mass_Misner_Sharp}, BGG only require the $K_{\theta\theta}$ component:
\begin{equation}
\label{eqB:extrinsic_curvature_K-theta}
    K_{\theta\theta} = \left(g_{\mu\nu} - s_\mu s_\nu \right) g^{\nu \gamma} \left(\partial_\gamma s_\theta - \Gamma^\sigma_{\gamma \nu} s_\sigma \right)
    = \frac{1}{2} s_\mu \partial^\mu g_{\theta\theta} \, ,
\end{equation}
where the Christoffel symbols are given by $\Gamma^\lambda_{\mu\nu} = g^{\lambda \delta}(\partial_\nu g_{\delta\mu} + \partial_\mu g_{\delta\nu} - \partial_\delta g_{\mu\nu})/2$. Before specializing to a FLRW manifold, we can eliminate dependence on $s_\mu$ by noting that (1) the proper time derivative for motion with $v^\mu$ is simply $\Dot{f} = v_\mu \partial^\mu f$ and (2) any four-vector may be decomposed into components parallel to $v^\mu$ and $s^\mu$. This decomposition for derivatives of the metric tensor looks like
\begin{equation}
    \partial_\mu g = - (v_\lambda \partial^\lambda g) v_\mu + \left(s_\nu \partial^\nu\right) s_\mu = -\Dot{g} v_\mu + \left(s_\nu \partial^\nu g\right) s_\mu \, .
\end{equation}
The squared norm of this expression,
\begin{equation}
    \left(\partial_\mu g\right) \left(\partial^\mu g\right) = -\Dot{g}^2 + (s_\lambda \partial^\lambda g)^2 \, ,
\end{equation}
allows us to rewrite Eq.~\eqref{eqB:extrinsic_curvature_K-theta} as
\begin{equation}
    K_{\theta\theta} = \frac{1}{2} \sqrt{\left(\Dot{g}_{\theta\theta}\right)^2 + \left(\partial_\mu g_{\theta\theta}\right) \left(\partial^\mu g_{\theta\theta}\right)} \, .
\end{equation}

The flat FLRW manifold assumed in our FOPT analyses has
\begin{equation}
    g_{\theta\theta} = a(t)^2 \chi^2
    \quad \implies \quad
    K_{\theta\theta} = r \sqrt{\Dot{r}^2 + 1 - H^2 r^2} \, ,
\end{equation}
where we evaluate the extrinsic curvature on the spherical domain wall with physical radius $r(\tau) \equiv a(t) \chi(\tau)$. The Friedmann equation $H^2 = 8\pi G (\rho_V + \rho_R)/3 \equiv H_V^2 + H_R^2$ reveals exactly how vacuum and radiation energies enter consideration. The BGG derivation has only a vacuum component, and therefore our FOPT analyses require
\begin{equation}
    H_R^2 r^2 \ll \left|\Dot{r}^2 + 1 - H_V^2 r^2\right| \, ,
\end{equation}
at times near PBH formation. Taking this right-hand side to be of order 1,
\begin{equation}
    r \ll 1 / H_R \, .
\end{equation}

\bibliography{main}

@article{LIGOScientific:2014pky,
    author = "Aasi, J. and others",
    collaboration = "LIGO Scientific",
    title = "{Advanced LIGO}",
    eprint = "1411.4547",
    archivePrefix = "arXiv",
    primaryClass = "gr-qc",
    doi = "10.1088/0264-9381/32/7/074001",
    journal = "Class. Quant. Grav.",
    volume = "32",
    pages = "074001",
    year = "2015"
}

@article{Shoemaker:2019bqt,
    author = "Shoemaker, David",
    collaboration = "LIGO Scientific",
    title = "{Gravitational wave astronomy with LIGO and similar detectors in the next decade}",
    eprint = "1904.03187",
    archivePrefix = "arXiv",
    primaryClass = "gr-qc",
    month = "4",
    year = "2019"
}

@article{Punturo:2010zz,
    author = "Punturo, M. and others",
    editor = "Ricci, Fulvio",
    title = "{The Einstein Telescope: A third-generation gravitational wave observatory}",
    doi = "10.1088/0264-9381/27/19/194002",
    journal = "Class. Quant. Grav.",
    volume = "27",
    pages = "194002",
    year = "2010"
}

@article{EPTA:2011kjn,
    author = "van Haasteren, R. and others",
    collaboration = "EPTA",
    title = "{Placing limits on the stochastic gravitational-wave background using European Pulsar Timing Array data}",
    eprint = "1103.0576",
    archivePrefix = "arXiv",
    primaryClass = "astro-ph.CO",
    doi = "10.1111/j.1365-2966.2011.18613.x",
    journal = "Mon. Not. Roy. Astron. Soc.",
    volume = "414",
    number = "4",
    pages = "3117--3128",
    year = "2011",
    note = "[Erratum: Mon.Not.Roy.Astron.Soc. 425, 1597 (2012)]"
}

@article{NANOGrav:2020bcs,
    author = "Arzoumanian, Zaven and others",
    collaboration = "NANOGrav",
    title = "{The NANOGrav 12.5 yr Data Set: Search for an Isotropic Stochastic Gravitational-wave Background}",
    eprint = "2009.04496",
    archivePrefix = "arXiv",
    primaryClass = "astro-ph.HE",
    doi = "10.3847/2041-8213/abd401",
    journal = "Astrophys. J. Lett.",
    volume = "905",
    number = "2",
    pages = "L34",
    year = "2020"
}

@article{Lazio:2017fos,
    author = "Lazio, T. Joseph W. and Bhaskaran, S. and Cutler, C. and Folkner, W. M. and Park, R. S. and Ellis, J. A. and Ely, T. and Taylor, S. R. and Vallisneri, M.",
    editor = "Sanidas, Sotiris and Weltevrede, Patrick and Preston, Lina Levin and Perera, Benetge B. P.",
    title = "{Solar System Ephemerides, Pulsar Timing, Gravitational Waves, \textbackslash{}\& Navigation}",
    eprint = "1801.02898",
    archivePrefix = "arXiv",
    primaryClass = "astro-ph.IM",
    doi = "10.1017/S1743921317009711",
    journal = "IAU Symp.",
    volume = "337",
    pages = "150--153",
    year = "2017"
}

@article{brown2018gaia,
  title={Gaia data release 2-summary of the contents and survey properties},
  author={Brown, AGA and Vallenari, Antonella and Prusti, TJDBJH and De Bruijne, JHJ and Babusiaux, C and Bailer-Jones, CAL and Biermann, M and Evans, Dafydd Wyn and Eyer, Laurent and Jansen, Femke and others},
  journal={Astronomy \& astrophysics},
  volume={616},
  pages={A1},
  year={2018},
  publisher={EDP sciences}
}

@ARTICLE{2018FrASS...5...11V,
       author = {{Vallenari}, Antonella},
        title = "{The Future of Astrometry in Space}",
      journal = {Frontiers in Astronomy and Space Sciences},
         year = 2018,
        month = apr,
       volume = {5},
          eid = {11},
        pages = {11},
          doi = {10.3389/fspas.2018.00011},
       adsurl = {https://ui.adsabs.harvard.edu/abs/2018FrASS...5...11V}
}

@article{Caprini:2019egz,
    author = "Caprini, Chiara and others",
    title = "{Detecting gravitational waves from cosmological phase transitions with LISA: an update}",
    eprint = "1910.13125",
    archivePrefix = "arXiv",
    primaryClass = "astro-ph.CO",
    reportNumber = "DESY-19-159, IPPP/19/27, HIP-2019-14/TH, MITP/19-066, IFT-UAM/CSIC-19-139",
    doi = "10.1088/1475-7516/2020/03/024",
    journal = "JCAP",
    volume = "03",
    pages = "024",
    year = "2020"
}

@article{amaro2017laser,
  title={Laser interferometer space antenna},
  author={Amaro-Seoane, Pau and Audley, Heather and Babak, Stanislav and Baker, John and Barausse, Enrico and Bender, Peter and Berti, Emanuele and Binetruy, Pierre and Born, Michael and Bortoluzzi, Daniele and others},
  journal={arXiv preprint arXiv:1702.00786},
  year={2017}
}

@article{Robson:2018ifk,
    author = "Robson, Travis and Cornish, Neil J. and Liu, Chang",
    title = "{The construction and use of LISA sensitivity curves}",
    eprint = "1803.01944",
    archivePrefix = "arXiv",
    primaryClass = "astro-ph.HE",
    doi = "10.1088/1361-6382/ab1101",
    journal = "Class. Quant. Grav.",
    volume = "36",
    number = "10",
    pages = "105011",
    year = "2019"
}

@article{Ruan:2018tsw,
    author = "Ruan, Wen-Hong and Guo, Zong-Kuan and Cai, Rong-Gen and Zhang, Yuan-Zhong",
    title = "{Taiji program: Gravitational-wave sources}",
    eprint = "1807.09495",
    archivePrefix = "arXiv",
    primaryClass = "gr-qc",
    doi = "10.1142/S0217751X2050075X",
    journal = "Int. J. Mod. Phys. A",
    volume = "35",
    number = "17",
    pages = "2050075",
    year = "2020"
}

@article{TianQin:2015yph,
    author = "Luo, Jun and others",
    collaboration = "TianQin",
    title = "{TianQin: a space-borne gravitational wave detector}",
    eprint = "1512.02076",
    archivePrefix = "arXiv",
    primaryClass = "astro-ph.IM",
    doi = "10.1088/0264-9381/33/3/035010",
    journal = "Class. Quant. Grav.",
    volume = "33",
    number = "3",
    pages = "035010",
    year = "2016"
}

@article{Gong:2014mca,
    author = "Gong, Xuefei and others",
    editor = "Ciani, Giacomo and Conklin, John W. and Mueller, Guido",
    title = "{Descope of the ALIA mission}",
    eprint = "1410.7296",
    archivePrefix = "arXiv",
    primaryClass = "gr-qc",
    doi = "10.1088/1742-6596/610/1/012011",
    journal = "J. Phys. Conf. Ser.",
    volume = "610",
    number = "1",
    pages = "012011",
    year = "2015"
}

@article{Corbin:2005ny,
    author = "Corbin, Vincent and Cornish, Neil J.",
    title = "{Detecting the cosmic gravitational wave background with the big bang observer}",
    eprint = "gr-qc/0512039",
    archivePrefix = "arXiv",
    doi = "10.1088/0264-9381/23/7/014",
    journal = "Class. Quant. Grav.",
    volume = "23",
    pages = "2435--2446",
    year = "2006"
}

@article{Yagi:2011wg,
    author = "Yagi, Kent and Seto, Naoki",
    title = "{Detector configuration of DECIGO/BBO and identification of cosmological neutron-star binaries}",
    eprint = "1101.3940",
    archivePrefix = "arXiv",
    primaryClass = "astro-ph.CO",
    doi = "10.1103/PhysRevD.83.044011",
    journal = "Phys. Rev. D",
    volume = "83",
    pages = "044011",
    year = "2011",
    note = "[Erratum: Phys.Rev.D 95, 109901 (2017)]"
}

@article{Kawamura:2020pcg,
    author = "Kawamura, Seiji and others",
    title = "{Current status of space gravitational wave antenna DECIGO and B-DECIGO}",
    eprint = "2006.13545",
    archivePrefix = "arXiv",
    primaryClass = "gr-qc",
    doi = "10.1093/ptep/ptab019",
    journal = "PTEP",
    volume = "2021",
    number = "5",
    pages = "05A105",
    year = "2021"
}

@article{Kawamura:2006up,
    author = "Kawamura, S. and others",
    editor = "Mio, N.",
    title = "{The Japanese space gravitational wave antenna DECIGO}",
    doi = "10.1088/0264-9381/23/8/S17",
    journal = "Class. Quant. Grav.",
    volume = "23",
    pages = "S125--S132",
    year = "2006"
}

@article{A+:LIGO,
    author = "T. L. S. Collaboration",
    title = "LIGO DCC-T1400316",
    year = "2014"
}

@article{LIGOScientific:2016wof,
    author = "Abbott, Benjamin P and others",
    collaboration = "LIGO Scientific",
    title = "{Exploring the Sensitivity of Next Generation Gravitational Wave Detectors}",
    eprint = "1607.08697",
    archivePrefix = "arXiv",
    primaryClass = "astro-ph.IM",
    reportNumber = "LIGO-P1600143",
    doi = "10.1088/1361-6382/aa51f4",
    journal = "Class. Quant. Grav.",
    volume = "34",
    number = "4",
    pages = "044001",
    year = "2017"
}

@article{Planck:2018vyg,
    author = "Aghanim, N. and others",
    collaboration = "Planck",
    title = "{Planck 2018 results. VI. Cosmological parameters}",
    eprint = "1807.06209",
    archivePrefix = "arXiv",
    primaryClass = "astro-ph.CO",
    doi = "10.1051/0004-6361/201833910",
    journal = "Astron. Astrophys.",
    volume = "641",
    pages = "A6",
    year = "2020",
    note = "[Erratum: Astron.Astrophys. 652, C4 (2021)]"
}

@article{ParticleDataGroup:2024cfk,
    author = "Navas, S. and others",
    collaboration = "Particle Data Group",
    title = "{Review of particle physics}",
    doi = "10.1103/PhysRevD.110.030001",
    journal = "Phys. Rev. D",
    volume = "110",
    number = "3",
    pages = "030001",
    year = "2024",
    url = {https://pdg.lbl.gov/}
}

@article{Carr:2020gox,
    author = "Carr, Bernard and Kohri, Kazunori and Sendouda, Yuuiti and Yokoyama, Jun'ichi",
    title = "{Constraints on primordial black holes}",
    eprint = "2002.12778",
    archivePrefix = "arXiv",
    primaryClass = "astro-ph.CO",
    reportNumber = "RESCEU-03/20; KEK-Cosmo-249; KEK-TH-2199; IPMU20-0024",
    doi = "10.1088/1361-6633/ac1e31",
    journal = "Rept. Prog. Phys.",
    volume = "84",
    number = "11",
    pages = "116902",
    year = "2021"
}

@article{Page:1976wx,
    author = "Page, Don N. and Hawking, S. W.",
    title = "{Gamma rays from primordial black holes}",
    doi = "10.1086/154350",
    journal = "Astrophys. J.",
    volume = "206",
    pages = "1--7",
    year = "1976"
}

@article{Carr:1998fw,
    author = "Carr, Bernard J. and MacGibbon, J. H.",
    editor = "Cline, D. B.",
    title = "{Cosmic rays from primordial black holes and constraints on the early universe}",
    doi = "10.1016/S0370-1573(98)00039-8",
    journal = "Phys. Rept.",
    volume = "307",
    pages = "141--154",
    year = "1998"
}

@article{Carr:2009jm,
    author = "Carr, B. J. and Kohri, Kazunori and Sendouda, Yuuiti and Yokoyama, Jun'ichi",
    title = "{New cosmological constraints on primordial black holes}",
    eprint = "0912.5297",
    archivePrefix = "arXiv",
    primaryClass = "astro-ph.CO",
    reportNumber = "RESCEU-31-09, TU-852, YITP-09-112",
    doi = "10.1103/PhysRevD.81.104019",
    journal = "Phys. Rev. D",
    volume = "81",
    pages = "104019",
    year = "2010"
}

@article{Smyth:2019whb,
    author = "Smyth, Nolan and Profumo, Stefano and English, Samuel and Jeltema, Tesla and McKinnon, Kevin and Guhathakurta, Puragra",
    title = "{Updated Constraints on Asteroid-Mass Primordial Black Holes as Dark Matter}",
    eprint = "1910.01285",
    archivePrefix = "arXiv",
    primaryClass = "astro-ph.CO",
    doi = "10.1103/PhysRevD.101.063005",
    journal = "Phys. Rev. D",
    volume = "101",
    number = "6",
    pages = "063005",
    year = "2020"
}

@article{Niikura:2019kqi,
    author = "Niikura, Hiroko and Takada, Masahiro and Yokoyama, Shuichiro and Sumi, Takahiro and Masaki, Shogo",
    title = "{Constraints on Earth-mass primordial black holes from OGLE 5-year microlensing events}",
    eprint = "1901.07120",
    archivePrefix = "arXiv",
    primaryClass = "astro-ph.CO",
    doi = "10.1103/PhysRevD.99.083503",
    journal = "Phys. Rev. D",
    volume = "99",
    number = "8",
    pages = "083503",
    year = "2019"
}

@article{EROS-2:2006ryy,
    author = "Tisserand, P. and others",
    collaboration = "EROS-2",
    title = "{Limits on the Macho Content of the Galactic Halo from the EROS-2 Survey of the Magellanic Clouds}",
    eprint = "astro-ph/0607207",
    archivePrefix = "arXiv",
    doi = "10.1051/0004-6361:20066017",
    journal = "Astron. Astrophys.",
    volume = "469",
    pages = "387--404",
    year = "2007"
}

@article{Bird:2022wvk,
    author = "Bird, Simeon and others",
    title = "{Snowmass2021 Cosmic Frontier White Paper: Primordial black hole dark matter}",
    eprint = "2203.08967",
    archivePrefix = "arXiv",
    primaryClass = "hep-ph",
    reportNumber = "FERMILAB-PUB-22-195-PPD",
    doi = "10.1016/j.dark.2023.101231",
    journal = "Phys. Dark Univ.",
    volume = "41",
    pages = "101231",
    year = "2023"
}

@misc{fardeen2024astrometric,
      title={Astrometric Microlensing by Primordial Black Holes with The Roman Space Telescope}, 
      author={James Fardeen and Peter McGill and Scott E. Perkins and William A. Dawson and Natasha S. Abrams and Jessica R. Lu and Ming-Feng Ho and Simeon Bird},
      year={2024},
      eprint={2312.13249},
      archivePrefix={arXiv},
      primaryClass={astro-ph.GA}
}

@misc{spergel2015widefield,
      title={Wide-Field InfrarRed Survey Telescope-Astrophysics Focused Telescope Assets WFIRST-AFTA 2015 Report}, 
      author={D. Spergel and others},
      year={2015},
      eprint={1503.03757},
      archivePrefix={arXiv},
      primaryClass={astro-ph.IM}
}

@article{AMEGO:2019gny,
    author = "Caputo, Regina and others",
    collaboration = "AMEGO",
    title = "{All-sky Medium Energy Gamma-ray Observatory: Exploring the Extreme Multimessenger Universe}",
    eprint = "1907.07558",
    archivePrefix = "arXiv",
    primaryClass = "astro-ph.IM",
    month = "7",
    year = "2019"
}

@article{Coogan:2020tuf,
    author = "Coogan, Adam and Morrison, Logan and Profumo, Stefano",
    title = "{Direct Detection of Hawking Radiation from Asteroid-Mass Primordial Black Holes}",
    eprint = "2010.04797",
    archivePrefix = "arXiv",
    primaryClass = "astro-ph.CO",
    doi = "10.1103/PhysRevLett.126.171101",
    journal = "Phys. Rev. Lett.",
    volume = "126",
    number = "17",
    pages = "171101",
    year = "2021"
}

@article{Linde:1977mm,
    author = "Linde, Andrei D.",
    title = "{On the Vacuum Instability and the Higgs Meson Mass}",
    reportNumber = "LEBEDEV-77-112",
    doi = "10.1016/0370-2693(77)90664-5",
    journal = "Phys. Lett. B",
    volume = "70",
    pages = "306--308",
    year = "1977"
}

@article{LINDE198137,
title = {Fate of the false vacuum at finite temperature: Theory and applications},
journal = {Physics Letters B},
volume = {100},
number = {1},
pages = {37-40},
year = {1981},
issn = {0370-2693},
doi = {https://doi.org/10.1016/0370-2693(81)90281-1},
url = {https://www.sciencedirect.com/science/article/pii/0370269381902811},
author = {A.D. Linde}
}

@inproceedings{Quiros:1999jp,
    author = "Quiros, Mariano",
    title = "{Finite temperature field theory and phase transitions}",
    booktitle = "{ICTP Summer School in High-Energy Physics and Cosmology}",
    eprint = "hep-ph/9901312",
    archivePrefix = "arXiv",
    reportNumber = "IEM-FT-187-99",
    pages = "187--259",
    month = "1",
    year = "1999"
}

@article{Marfatia:2021hcp,
    author = "Marfatia, Danny and Tseng, Po-Yan",
    title = "{Correlated signals of first-order phase transitions and primordial black hole evaporation}",
    eprint = "2112.14588",
    archivePrefix = "arXiv",
    primaryClass = "hep-ph",
    doi = "10.1007/JHEP08(2022)001",
    journal = "JHEP",
    volume = "08",
    pages = "001",
    year = "2022",
    note = "[Erratum: JHEP 08, 249 (2022)]"
}

@article{Gehrman:2023esa,
    author = "Gehrman, Thomas C. and Shams Es Haghi, Barmak and Sinha, Kuver and Xu, Tao",
    title = "{The primordial black holes that disappeared: connections to dark matter and MHz-GHz gravitational Waves}",
    eprint = "2304.09194",
    archivePrefix = "arXiv",
    primaryClass = "hep-ph",
    reportNumber = "UTWI-10-2023",
    doi = "10.1088/1475-7516/2023/10/001",
    journal = "JCAP",
    volume = "10",
    pages = "001",
    year = "2023"
}

@article{Lewicki:2023mik,
    author = {Lewicki, Marek and M\"u\"ursepp, Kristjan and Pata, Joosep and Vasar, Martin and Vaskonen, Ville and Veerm\"ae, Hardi},
    title = "{Dynamics of false vacuum bubbles with trapped particles}",
    eprint = "2305.07702",
    archivePrefix = "arXiv",
    primaryClass = "hep-ph",
    doi = "10.1103/PhysRevD.108.036023",
    journal = "Phys. Rev. D",
    volume = "108",
    number = "3",
    pages = "036023",
    year = "2023"
}

@article{Iso:2009ss,
    author = "Iso, Satoshi and Okada, Nobuchika and Orikasa, Yuta",
    title = "{Classically conformal $B^-$ L extended Standard Model}",
    eprint = "0902.4050",
    archivePrefix = "arXiv",
    primaryClass = "hep-ph",
    reportNumber = "KEK-TH-1303",
    doi = "10.1016/j.physletb.2009.04.046",
    journal = "Phys. Lett. B",
    volume = "676",
    pages = "81--87",
    year = "2009"
}

@article{Iso:2009nw,
    author = "Iso, Satoshi and Okada, Nobuchika and Orikasa, Yuta",
    title = "{The minimal B-L model naturally realized at TeV scale}",
    eprint = "0909.0128",
    archivePrefix = "arXiv",
    primaryClass = "hep-ph",
    reportNumber = "KEK-TH-1327",
    doi = "10.1103/PhysRevD.80.115007",
    journal = "Phys. Rev. D",
    volume = "80",
    pages = "115007",
    year = "2009"
}

@article{Jinno:2016knw,
    author = "Jinno, Ryusuke and Takimoto, Masahiro",
    title = "{Probing a classically conformal B-L model with gravitational waves}",
    eprint = "1604.05035",
    archivePrefix = "arXiv",
    primaryClass = "hep-ph",
    reportNumber = "KEK-TH-1896",
    doi = "10.1103/PhysRevD.95.015020",
    journal = "Phys. Rev. D",
    volume = "95",
    number = "1",
    pages = "015020",
    year = "2017"
}

@article{Marzo:2018nov,
    author = "Marzo, Carlo and Marzola, Luca and Vaskonen, Ville",
    title = "{Phase transition and vacuum stability in the classically conformal B\textendash{}L model}",
    eprint = "1811.11169",
    archivePrefix = "arXiv",
    primaryClass = "hep-ph",
    reportNumber = "KCL-PH-TH/2018-68",
    doi = "10.1140/epjc/s10052-019-7076-x",
    journal = "Eur. Phys. J. C",
    volume = "79",
    number = "7",
    pages = "601",
    year = "2019"
}

@article{Lewicki:2023ioy,
    author = "Lewicki, Marek and Toczek, Piotr and Vaskonen, Ville",
    title = "{Primordial black holes from strong first-order phase transitions}",
    eprint = "2305.04924",
    archivePrefix = "arXiv",
    primaryClass = "astro-ph.CO",
    doi = "10.1007/JHEP09(2023)092",
    journal = "JHEP",
    volume = "09",
    pages = "092",
    year = "2023"
}

@article{Ellis:2020nnr,
    author = "Ellis, John and Lewicki, Marek and Vaskonen, Ville",
    title = "{Updated predictions for gravitational waves produced in a strongly supercooled phase transition}",
    eprint = "2007.15586",
    archivePrefix = "arXiv",
    primaryClass = "astro-ph.CO",
    reportNumber = "KCL-PH-TH/2020-40, CERN-TH-2020-129",
    doi = "10.1088/1475-7516/2020/11/020",
    journal = "JCAP",
    volume = "11",
    pages = "020",
    year = "2020"
}

@article{Buchmuller:1991ce,
    author = "Buchmuller, W. and Greub, C. and Minkowski, P.",
    title = "{Neutrino masses, neutral vector bosons and the scale of B-L breaking}",
    reportNumber = "DESY-91-053",
    doi = "10.1016/0370-2693(91)90952-M",
    journal = "Phys. Lett. B",
    volume = "267",
    pages = "395--399",
    year = "1991"
}

@article{Babu:1989ex,
    author = "Babu, K. S. and Mohapatra, Rabindra N.",
    title = "{Quantization of Electric Charge From Anomaly Constraints and a Majorana Neutrino}",
    reportNumber = "MDDP-PP-90-011",
    doi = "10.1103/PhysRevD.41.271",
    journal = "Phys. Rev. D",
    volume = "41",
    pages = "271",
    year = "1990"
}

@article{Sher:1988mj,
    author = "Sher, Marc",
    title = "{Electroweak Higgs Potentials and Vacuum Stability}",
    reportNumber = "WU-TH-88-8",
    doi = "10.1016/0370-1573(89)90061-6",
    journal = "Phys. Rept.",
    volume = "179",
    pages = "273--418",
    year = "1989"
}

@article{Meissner:2008uw,
    author = "Meissner, Krzysztof A. and Nicolai, Hermann",
    title = "{Renormalization Group and Effective Potential in Classically Conformal Theories}",
    eprint = "0809.1338",
    archivePrefix = "arXiv",
    primaryClass = "hep-th",
    reportNumber = "AEI-2008-062, LPTENS-08-50",
    journal = "Acta Phys. Polon. B",
    volume = "40",
    pages = "2737--2752",
    year = "2009"
}

@article{Coleman:1973jx,
    author = "Coleman, Sidney R. and Weinberg, Erick J.",
    title = "{Radiative Corrections as the Origin of Spontaneous Symmetry Breaking}",
    doi = "10.1103/PhysRevD.7.1888",
    journal = "Phys. Rev. D",
    volume = "7",
    pages = "1888--1910",
    year = "1973"
}

@article{Maggiore:1999vm,
    author = "Maggiore, Michele",
    title = "{Gravitational wave experiments and early universe cosmology}",
    eprint = "gr-qc/9909001",
    archivePrefix = "arXiv",
    reportNumber = "IFUP-TH-20-99",
    doi = "10.1016/S0370-1573(99)00102-7",
    journal = "Phys. Rept.",
    volume = "331",
    pages = "283--367",
    year = "2000"
}

@article{PhysRevD.59.102001,
  title = {Detecting a stochastic background of gravitational radiation: Signal processing strategies and sensitivities},
  author = {Allen, Bruce and Romano, Joseph D.},
  journal = {Phys. Rev. D},
  volume = {59},
  issue = {10},
  pages = {102001},
  numpages = {41},
  year = {1999},
  month = {Mar},
  publisher = {American Physical Society},
  doi = {10.1103/PhysRevD.59.102001},
  url = {https://link.aps.org/doi/10.1103/PhysRevD.59.102001}
}

@article{Caprini:2015zlo,
    author = "Caprini, Chiara and others",
    title = "{Science with the space-based interferometer eLISA. II: Gravitational waves from cosmological phase transitions}",
    eprint = "1512.06239",
    archivePrefix = "arXiv",
    primaryClass = "astro-ph.CO",
    reportNumber = "DESY-15-246",
    doi = "10.1088/1475-7516/2016/04/001",
    journal = "JCAP",
    volume = "04",
    pages = "001",
    year = "2016"
}

@article{Coleman:1977py,
    author = "Coleman, Sidney R.",
    title = "{The Fate of the False Vacuum. 1. Semiclassical Theory}",
    reportNumber = "HUTP-77-A004",
    doi = "10.1103/PhysRevD.16.1248",
    journal = "Phys. Rev. D",
    volume = "15",
    pages = "2929--2936",
    year = "1977",
    note = "[Erratum: Phys.Rev.D 16, 1248 (1977)]"
}

@article{Linde:1981zj,
    author = "Linde, Andrei D.",
    title = "{Decay of the False Vacuum at Finite Temperature}",
    reportNumber = "LEBEDEV-81-265",
    doi = "10.1016/0550-3213(83)90072-X",
    journal = "Nucl. Phys. B",
    volume = "216",
    pages = "421",
    year = "1983",
    note = "[Erratum: Nucl.Phys.B 223, 544 (1983)]"
}

@article{Kierkla:2022odc,
    author = "Kierkla, Maciej and Karam, Alexandros and Swiezewska, Bogumila",
    title = "{Conformal model for gravitational waves and dark matter: a status update}",
    eprint = "2210.07075",
    archivePrefix = "arXiv",
    primaryClass = "astro-ph.CO",
    doi = "10.1007/JHEP03(2023)007",
    journal = "JHEP",
    volume = "03",
    pages = "007",
    year = "2023"
}

@article{Costa:2025pew,
    author = "Costa, Francesco and Hoefken Zink, Jaime and Lucente, Michele and Pascoli, Silvia and Rosauro-Alcaraz, Salvador",
    title = "{ELENA: a software for fast and precise computation of first order phase transitions and gravitational waves production in particle physics models}",
    eprint = "2510.00289",
    archivePrefix = "arXiv",
    primaryClass = "hep-ph",
    month = "9",
    year = "2025"
}

@article{Espinosa:2018hue,
    author = "Espinosa, J. R.",
    title = "{A Fresh Look at the Calculation of Tunneling Actions}",
    eprint = "1805.03680",
    archivePrefix = "arXiv",
    primaryClass = "hep-th",
    doi = "10.1088/1475-7516/2018/07/036",
    journal = "JCAP",
    volume = "07",
    pages = "036",
    year = "2018"
}

@article{Espinosa:2010hh,
    author = "Espinosa, Jose R. and Konstandin, Thomas and No, Jose M. and Servant, Geraldine",
    title = "{Energy Budget of Cosmological First-order Phase Transitions}",
    eprint = "1004.4187",
    archivePrefix = "arXiv",
    primaryClass = "hep-ph",
    reportNumber = "CERN-PH-TH-2010-027",
    doi = "10.1088/1475-7516/2010/06/028",
    journal = "JCAP",
    volume = "06",
    pages = "028",
    year = "2010"
}

@article{Bodeker:2009qy,
    author = "Bodeker, Dietrich and Moore, Guy D.",
    title = "{Can electroweak bubble walls run away?}",
    eprint = "0903.4099",
    archivePrefix = "arXiv",
    primaryClass = "hep-ph",
    doi = "10.1088/1475-7516/2009/05/009",
    journal = "JCAP",
    volume = "05",
    pages = "009",
    year = "2009"
}

@article{Bodeker:2017cim,
    author = "Bodeker, Dietrich and Moore, Guy D.",
    title = "{Electroweak Bubble Wall Speed Limit}",
    eprint = "1703.08215",
    archivePrefix = "arXiv",
    primaryClass = "hep-ph",
    doi = "10.1088/1475-7516/2017/05/025",
    journal = "JCAP",
    volume = "05",
    pages = "025",
    year = "2017"
}

@article{Gouttenoire:2021kjv,
    author = "Gouttenoire, Yann and Jinno, Ryusuke and Sala, Filippo",
    title = "{Friction pressure on relativistic bubble walls}",
    eprint = "2112.07686",
    archivePrefix = "arXiv",
    primaryClass = "hep-ph",
    reportNumber = "DESY-21-147, IFT-UAM/CSIC-21-146",
    doi = "10.1007/JHEP05(2022)004",
    journal = "JHEP",
    volume = "05",
    pages = "004",
    year = "2022"
}

@article{10.1063/1.1338506,
    author = {Lorenz, Christian D. and Ziff, Robert M.},
    title = {Precise determination of the critical percolation threshold for the three-dimensional “Swiss cheese” model using a growth algorithm},
    journal = {The Journal of Chemical Physics},
    volume = {114},
    number = {8},
    pages = {3659-3661},
    year = {2001},
    month = {02},
    issn = {0021-9606},
    doi = {10.1063/1.1338506},
    url = {https://doi.org/10.1063/1.1338506},
}

@article{LIN2018299,
title = {Continuum percolation of porous media via random packing of overlapping cube-like particles},
journal = {Theoretical and Applied Mechanics Letters},
volume = {8},
number = {5},
pages = {299-303},
year = {2018},
issn = {2095-0349},
doi = {https://doi.org/10.1016/j.taml.2018.05.007},
url = {https://www.sciencedirect.com/science/article/pii/S209503491830196X},
author = {Jianjun Lin and Huisu Chen}
}

@article{LI2020112815,
title = {Numerical study for the percolation threshold and transport properties of porous composites comprising non-centrosymmetrical superovoidal pores},
journal = {Computer Methods in Applied Mechanics and Engineering},
volume = {361},
pages = {112815},
year = {2020},
issn = {0045-7825},
doi = {https://doi.org/10.1016/j.cma.2019.112815},
url = {https://www.sciencedirect.com/science/article/pii/S0045782519307078},
author = {Mingqi Li and Huisu Chen and Jianjun Lin}
}

@article{Guth:1979bh,
    author = "Guth, Alan H. and Tye, S. H. H.",
    title = "{Phase Transitions and Magnetic Monopole Production in the Very Early Universe}",
    reportNumber = "SLAC-PUB-2448, CLNS-79-441",
    doi = "10.1103/PhysRevLett.44.631",
    journal = "Phys. Rev. Lett.",
    volume = "44",
    pages = "631",
    year = "1980",
    note = "[Erratum: Phys.Rev.Lett. 44, 963 (1980)]"
}

@article{Guth:1981uk,
    author = "Guth, Alan H. and Weinberg, Erick J.",
    title = "{Cosmological Consequences of a First Order Phase Transition in the SU(5) Grand Unified Model}",
    reportNumber = "CU-TP-183",
    doi = "10.1103/PhysRevD.23.876",
    journal = "Phys. Rev. D",
    volume = "23",
    pages = "876",
    year = "1981"
}

@article{Dent:2025bwo,
    author = "Dent, James B. and Dutta, Bhaskar and Rai, Mudit",
    title = "{Primordial black holes at the junction}",
    eprint = "2510.05236",
    archivePrefix = "arXiv",
    primaryClass = "hep-ph",
    doi = "10.1007/JHEP05(2026)279",
    journal = "JHEP",
    volume = "05",
    pages = "279",
    year = "2026"
}

@article{Flores:2024lng,
    author = "Flores, Marcos M. and Kusenko, Alexander and Sasaki, Misao",
    title = "{Revisiting formation of primordial black holes in a supercooled first-order phase transition}",
    eprint = "2402.13341",
    archivePrefix = "arXiv",
    primaryClass = "hep-ph",
    reportNumber = "IPMU23-0053, YITP-23-169",
    doi = "10.1103/PhysRevD.110.015005",
    journal = "Phys. Rev. D",
    volume = "110",
    number = "1",
    pages = "015005",
    year = "2024"
}

@article{Blau:1986cw,
    author = "Blau, Steven K. and Guendelman, E. I. and Guth, Alan H.",
    title = "{The Dynamics of False Vacuum Bubbles}",
    reportNumber = "MIT-CTP-1292",
    doi = "10.1103/PhysRevD.35.1747",
    journal = "Phys. Rev. D",
    volume = "35",
    pages = "1747",
    year = "1987"
}

@article{Enqvist:1991xw,
    author = "Enqvist, K. and Ignatius, J. and Kajantie, K. and Rummukainen, K.",
    title = "{Nucleation and bubble growth in a first order cosmological electroweak phase transition}",
    reportNumber = "HU-TFT-91-35",
    doi = "10.1103/PhysRevD.45.3415",
    journal = "Phys. Rev. D",
    volume = "45",
    pages = "3415--3428",
    year = "1992"
}

@article{Gross:2021qgx,
    author = "Gross, Christian and Landini, Giacomo and Strumia, Alessandro and Teresi, Daniele",
    title = "{Dark Matter as dark dwarfs and other macroscopic objects: multiverse relics?}",
    eprint = "2105.02840",
    archivePrefix = "arXiv",
    primaryClass = "hep-ph",
    doi = "10.1007/JHEP09(2021)033",
    journal = "JHEP",
    volume = "09",
    pages = "033",
    year = "2021"
}

@article{Baker:2021nyl,
    author = "Baker, Michael J. and Breitbach, Moritz and Kopp, Joachim and Mittnacht, Lukas",
    title = "{Primordial Black Holes from First-Order Cosmological Phase Transitions}",
    eprint = "2105.07481",
    archivePrefix = "arXiv",
    primaryClass = "astro-ph.CO",
    reportNumber = "CERN-TH-2021-079, MITP-21-023",
    month = "5",
    year = "2021"
}

@article{Baker:2021sno,
    author = "Baker, Michael J. and Breitbach, Moritz and Kopp, Joachim and Mittnacht, Lukas",
    title = "{Detailed Calculation of Primordial Black Hole Formation During First-Order Cosmological Phase Transitions}",
    eprint = "2110.00005",
    archivePrefix = "arXiv",
    primaryClass = "astro-ph.CO",
    reportNumber = "CERN-TH-2021-113, MITP-21-036",
    month = "9",
    year = "2021"
}

@article{Kawana:2021tde,
    author = "Kawana, Kiyoharu and Xie, Ke-Pan",
    title = "{Primordial black holes from a cosmic phase transition: The collapse of Fermi-balls}",
    eprint = "2106.00111",
    archivePrefix = "arXiv",
    primaryClass = "astro-ph.CO",
    doi = "10.1016/j.physletb.2021.136791",
    journal = "Phys. Lett. B",
    volume = "824",
    pages = "136791",
    year = "2022"
}

@article{Hong:2020est,
    author = "Hong, Jeong-Pyong and Jung, Sunghoon and Xie, Ke-Pan",
    title = "{Fermi-ball dark matter from a first-order phase transition}",
    eprint = "2008.04430",
    archivePrefix = "arXiv",
    primaryClass = "hep-ph",
    doi = "10.1103/PhysRevD.102.075028",
    journal = "Phys. Rev. D",
    volume = "102",
    number = "7",
    pages = "075028",
    year = "2020"
}

@article{Dent:2025lwe,
    author = "Dent, James B. and Dutta, Bhaskar and Kumar, Jason and Marfatia, Danny",
    title = "{Primordial black holes from Q-balls produced in a first-order phase transition}",
    eprint = "2505.21830",
    archivePrefix = "arXiv",
    primaryClass = "hep-ph",
    month = "5",
    year = "2025"
}

@article{Carr:1975qj,
    author = "Carr, Bernard J.",
    title = "{The Primordial black hole mass spectrum}",
    doi = "10.1086/153853",
    journal = "Astrophys. J.",
    volume = "201",
    pages = "1--19",
    year = "1975"
}

@article{Harada:2013epa,
    author = "Harada, Tomohiro and Yoo, Chul-Moon and Kohri, Kazunori",
    title = "{Threshold of primordial black hole formation}",
    eprint = "1309.4201",
    archivePrefix = "arXiv",
    primaryClass = "astro-ph.CO",
    reportNumber = "RUP-13-9, KEK-COSMO-129, KEK-TH-1668",
    doi = "10.1103/PhysRevD.88.084051",
    journal = "Phys. Rev. D",
    volume = "88",
    number = "8",
    pages = "084051",
    year = "2013",
    note = "[Erratum: Phys.Rev.D 89, 029903 (2014)]"
}

@article{Escriva:2021aeh,
    author = "Escriv{\`a}, Albert",
    title = "{PBH Formation from Spherically Symmetric Hydrodynamical Perturbations: A Review}",
    eprint = "2111.12693",
    archivePrefix = "arXiv",
    primaryClass = "gr-qc",
    doi = "10.3390/universe8020066",
    journal = "Universe",
    volume = "8",
    number = "2",
    pages = "66",
    year = "2022"
}

@article{Lu:2022paj,
    author = "Lu, Philip and Kawana, Kiyoharu and Xie, Ke-Pan",
    title = "{Old phase remnants in first-order phase transitions}",
    eprint = "2202.03439",
    archivePrefix = "arXiv",
    primaryClass = "astro-ph.CO",
    doi = "10.1103/PhysRevD.105.123503",
    journal = "Phys. Rev. D",
    volume = "105",
    number = "12",
    pages = "123503",
    year = "2022"
}

@article{Hawking:1975vcx,
    author = "Hawking, S. W.",
    editor = "Gibbons, G. W. and Hawking, S. W.",
    title = "{Particle Creation by Black Holes}",
    doi = "10.1007/BF02345020",
    journal = "Commun. Math. Phys.",
    volume = "43",
    pages = "199--220",
    year = "1975",
    note = "[Erratum: Commun.Math.Phys. 46, 206 (1976)]"
}

@article{Gell-Mann:1979vob,
    author = "Gell-Mann, Murray and Ramond, Pierre and Slansky, Richard",
    title = "{Complex Spinors and Unified Theories}",
    eprint = "1306.4669",
    archivePrefix = "arXiv",
    primaryClass = "hep-th",
    reportNumber = "PRINT-80-0576",
    journal = "Conf. Proc. C",
    volume = "790927",
    pages = "315--321",
    year = "1979"
}

@article{Mohapatra:1979ia,
    author = "Mohapatra, Rabindra N. and Senjanovic, Goran",
    title = "{Neutrino Mass and Spontaneous Parity Nonconservation}",
    reportNumber = "MDDP-TR-80-060, MDDP-PP-80-105, CCNY-HEP-79-10",
    doi = "10.1103/PhysRevLett.44.912",
    journal = "Phys. Rev. Lett.",
    volume = "44",
    pages = "912",
    year = "1980"
}

@article{Minkowski:1977sc,
    author = "Minkowski, Peter",
    title = "{$\mu \to e\gamma$ at a Rate of One Out of $10^{9}$ Muon Decays?}",
    reportNumber = "Print-77-0182 (BERN)",
    doi = "10.1016/0370-2693(77)90435-X",
    journal = "Phys. Lett. B",
    volume = "67",
    pages = "421--428",
    year = "1977"
}

@article{Yanagida:1979as,
    author = "Yanagida, Tsutomu",
    editor = "Sawada, Osamu and Sugamoto, Akio",
    title = "{Horizontal gauge symmetry and masses of neutrinos}",
    reportNumber = "KEK-79-18-95",
    journal = "Conf. Proc. C",
    volume = "7902131",
    pages = "95--99",
    year = "1979"
}

@article{Borsanyi:2016ksw,
    author = "Borsanyi, Sz. and others",
    title = "{Calculation of the axion mass based on high-temperature lattice quantum chromodynamics}",
    eprint = "1606.07494",
    archivePrefix = "arXiv",
    primaryClass = "hep-lat",
    reportNumber = "DESY-16-105",
    doi = "10.1038/nature20115",
    journal = "Nature",
    volume = "539",
    number = "7627",
    pages = "69--71",
    year = "2016"
}

@article{Giese:2020rtr,
    author = "Giese, Felix and Konstandin, Thomas and van de Vis, Jorinde",
    title = "{Model-independent energy budget of cosmological first-order phase transitions{\textemdash}A sound argument to go beyond the bag model}",
    eprint = "2004.06995",
    archivePrefix = "arXiv",
    primaryClass = "astro-ph.CO",
    reportNumber = "DESY-20-064",
    doi = "10.1088/1475-7516/2020/07/057",
    journal = "JCAP",
    volume = "07",
    number = "07",
    pages = "057",
    year = "2020"
}

@article{Ellis:2019oqb,
    author = "Ellis, John and Lewicki, Marek and No, Jos{\'e} Miguel and Vaskonen, Ville",
    title = "{Gravitational wave energy budget in strongly supercooled phase transitions}",
    eprint = "1903.09642",
    archivePrefix = "arXiv",
    primaryClass = "hep-ph",
    reportNumber = "KCL-PH-TH/2019-32, CERN-TH-2019-032, IFT-UAM/CSIC-19-32",
    doi = "10.1088/1475-7516/2019/06/024",
    journal = "JCAP",
    volume = "06",
    pages = "024",
    year = "2019"
}

@article{Athron:2023xlk,
    author = "Athron, Peter and Bal{\'a}zs, Csaba and Fowlie, Andrew and Morris, Lachlan and Wu, Lei",
    title = "{Cosmological phase transitions: From perturbative particle physics to gravitational waves}",
    eprint = "2305.02357",
    archivePrefix = "arXiv",
    primaryClass = "hep-ph",
    doi = "10.1016/j.ppnp.2023.104094",
    journal = "Prog. Part. Nucl. Phys.",
    volume = "135",
    pages = "104094",
    year = "2024"
}

@article{Murai:2025hse,
    author = "Murai, Kai and Sakurai, Kodai and Takahashi, Fuminobu",
    title = "{Primordial black hole formation via inverted bubble collapse}",
    eprint = "2502.02291",
    archivePrefix = "arXiv",
    primaryClass = "astro-ph.CO",
    reportNumber = "TU-1254",
    doi = "10.1007/JHEP07(2025)065",
    journal = "JHEP",
    volume = "07",
    pages = "065",
    year = "2025"
}

@article{Hashino:2025fse,
    author = "Hashino, Katsuya and Kanemura, Shinya and Takahashi, Tomo and Tanaka, Masanori and Yoo, Chul-Moon",
    title = "{Super-critical primordial black hole formation via delayed first-order electroweak phase transition}",
    eprint = "2501.11040",
    archivePrefix = "arXiv",
    primaryClass = "hep-ph",
    reportNumber = "OU-HET-1260",
    doi = "10.1088/1475-7516/2025/09/006",
    journal = "JCAP",
    volume = "09",
    pages = "006",
    year = "2025"
}

@article{Einhorn:1983fc,
    author = "Einhorn, Martin B. and Jones, D. R. Timothy",
    title = "{A NEW RENORMALIZATION GROUP APPROACH TO MULTISCALE PROBLEMS}",
    reportNumber = "UM-TH-83-16",
    doi = "10.1016/0550-3213(84)90127-5",
    journal = "Nucl. Phys. B",
    volume = "230",
    pages = "261--272",
    year = "1984"
}

@article{Ford:1992mv,
    author = "Ford, C. and Jones, D. R. T. and Stephenson, P. W. and Einhorn, M. B.",
    title = "{The Effective potential and the renormalization group}",
    eprint = "hep-lat/9210033",
    archivePrefix = "arXiv",
    reportNumber = "LTH-288, UM-TH-92-21",
    doi = "10.1016/0550-3213(93)90206-5",
    journal = "Nucl. Phys. B",
    volume = "395",
    pages = "17--34",
    year = "1993"
}

@article{Bando:1992wy,
    author = "Bando, Masako and Kugo, Taichiro and Maekawa, Nobuhiro and Nakano, Hiroaki",
    title = "{Improving the effective potential: Multimass scale case}",
    eprint = "hep-ph/9210229",
    archivePrefix = "arXiv",
    reportNumber = "KUNS-1162, KUNS1162",
    doi = "10.1143/PTP.90.405",
    journal = "Prog. Theor. Phys.",
    volume = "90",
    pages = "405--418",
    year = "1993"
}

@article{Ford:1994dt,
    author = "Ford, Christopher",
    title = "{Multiscale renormalization group improvement of the effective potential}",
    eprint = "hep-th/9404085",
    archivePrefix = "arXiv",
    reportNumber = "DIAS-STP-94-09",
    doi = "10.1103/PhysRevD.50.7531",
    journal = "Phys. Rev. D",
    volume = "50",
    pages = "7531--7537",
    year = "1994"
}

@article{Ford:1996hd,
    author = "Ford, C. and Wiesendanger, C.",
    title = "{A Multiscale subtraction scheme and partial renormalization group equations in the O(N) symmetric phi**4 theory}",
    eprint = "hep-ph/9604392",
    archivePrefix = "arXiv",
    reportNumber = "DIAS-STP-96-10",
    doi = "10.1103/PhysRevD.55.2202",
    journal = "Phys. Rev. D",
    volume = "55",
    pages = "2202--2217",
    year = "1997"
}

@article{Casas:1998cf,
    author = "Casas, J. A. and Di Clemente, V. and Quiros, M.",
    title = "{The Effective potential in the presence of several mass scales}",
    eprint = "hep-ph/9809275",
    archivePrefix = "arXiv",
    reportNumber = "IEM-FT-181-98",
    doi = "10.1016/S0550-3213(99)00262-X",
    journal = "Nucl. Phys. B",
    volume = "553",
    pages = "511--530",
    year = "1999"
}

@article{Steele:2014dsa,
    author = "Steele, T. G. and Wang, Zhi-Wei and McKeon, D. G. C.",
    title = "{Multiscale renormalization group methods for effective potentials with multiple scalar fields}",
    eprint = "1409.3489",
    archivePrefix = "arXiv",
    primaryClass = "hep-ph",
    doi = "10.1103/PhysRevD.90.105012",
    journal = "Phys. Rev. D",
    volume = "90",
    number = "10",
    pages = "105012",
    year = "2014"
}

@article{Chataignier:2018aud,
    author = "Chataignier, Leonardo and Prokopec, Tomislav and Schmidt, Michael G. and Swiezewska, Bogumila",
    title = "{Single-scale Renormalisation Group Improvement of Multi-scale Effective Potentials}",
    eprint = "1801.05258",
    archivePrefix = "arXiv",
    primaryClass = "hep-ph",
    doi = "10.1007/JHEP03(2018)014",
    journal = "JHEP",
    volume = "03",
    pages = "014",
    year = "2018"
}

@article{Leitao:2015fmj,
    author = "Leitao, Leonardo and Megevand, Ariel",
    title = "{Gravitational waves from a very strong electroweak phase transition}",
    eprint = "1512.08962",
    archivePrefix = "arXiv",
    primaryClass = "astro-ph.CO",
    doi = "10.1088/1475-7516/2016/05/037",
    journal = "JCAP",
    volume = "05",
    pages = "037",
    year = "2016"
}

@article{Ghoshal:2025iil,
    author = "Ghoshal, Anish and Spalding, Angus and White, Graham",
    title = "{Cosmic Strings Gravitational Wave Probe of Leptogenesis: Thermal, Non-thermal, Near-resonant and Flavourful}",
    eprint = "2512.14684",
    archivePrefix = "arXiv",
    primaryClass = "astro-ph.CO",
    month = "12",
    year = "2025"
}

@article{DeRomeri:2024iaw,
    author = "De Romeri, Valentina and Papoulias, Dimitrios K. and Ternes, Christoph A.",
    title = "{Bounds on new neutrino interactions from the first CE{\ensuremath{\nu}}NS data at direct detection experiments}",
    eprint = "2411.11749",
    archivePrefix = "arXiv",
    primaryClass = "hep-ph",
    doi = "10.1088/1475-7516/2025/05/012",
    journal = "JCAP",
    volume = "05",
    pages = "012",
    year = "2025"
}

@article{Bandyopadhyay:2018cwu,
    author = "Bandyopadhyay, Triparno and Bhattacharyya, Gautam and Das, Dipankar and Raychaudhuri, Amitava",
    title = "{Reappraisal of constraints on $Z^\prime$ models from unitarity and direct searches at the LHC}",
    eprint = "1803.07989",
    archivePrefix = "arXiv",
    primaryClass = "hep-ph",
    doi = "10.1103/PhysRevD.98.035027",
    journal = "Phys. Rev. D",
    volume = "98",
    number = "3",
    pages = "035027",
    year = "2018"
}

@article{Wang:2024gvt,
    author = "Wang, X. G. and Hunt-Smith, N. T. and Melnitchouk, W. and Sato, N. and Thomas, A. W.",
    collaboration = "[JAM collaboration (BSM Analysis Group){\,}]",
    title = "{Constraints on the U(1)B-L model from global QCD analysis}",
    eprint = "2410.01205",
    archivePrefix = "arXiv",
    primaryClass = "hep-ph",
    reportNumber = "ADP-24-14/T1253, JLAB-THY-24-4206",
    doi = "10.1103/PhysRevD.111.015019",
    journal = "Phys. Rev. D",
    volume = "111",
    number = "1",
    pages = "015019",
    year = "2025"
}

@article{Deppisch2019SearchingFA,
  title={Searching for a light $Z^\prime$ through Higgs production at the LHC},
  author={Frank F. Deppisch and Suchita Kulkarni and Wei Liu},
  journal={Physical Review D},
  year={2019},
  url={https://api.semanticscholar.org/CorpusID:201698287}
}

@article{Xie:2023cwi,
    author = "Xie, Ke-Pan",
    title = "{Pinning down the primordial black hole formation mechanism with gamma-rays and gravitational waves}",
    eprint = "2301.02352",
    archivePrefix = "arXiv",
    primaryClass = "astro-ph.CO",
    doi = "10.1088/1475-7516/2023/06/008",
    journal = "JCAP",
    volume = "06",
    pages = "008",
    year = "2023"
}

@article{Huang:2025hos,
    author = "Huang, Fa Peng and Idegawa, Chikako and Yang, Aidi",
    title = "{Primordial black hole formation and multimessenger signals in a complex singlet extension of the standard model}",
    eprint = "2510.24007",
    archivePrefix = "arXiv",
    primaryClass = "hep-ph",
    doi = "10.1103/cxh3-bpl6",
    journal = "Phys. Rev. D",
    volume = "113",
    number = "5",
    pages = "055013",
    year = "2026"
}

@article{Huang:2022him,
    author = "Huang, Peisi and Xie, Ke-Pan",
    title = "{Primordial black holes from an electroweak phase transition}",
    eprint = "2201.07243",
    archivePrefix = "arXiv",
    primaryClass = "hep-ph",
    doi = "10.1103/PhysRevD.105.115033",
    journal = "Phys. Rev. D",
    volume = "105",
    number = "11",
    pages = "115033",
    year = "2022"
}

@article{Lewicki:2024ghw,
    author = "Lewicki, Marek and Toczek, Piotr and Vaskonen, Ville",
    title = "{Black Holes and Gravitational Waves from Slow First-Order Phase Transitions}",
    eprint = "2402.04158",
    archivePrefix = "arXiv",
    primaryClass = "astro-ph.CO",
    doi = "10.1103/PhysRevLett.133.221003",
    journal = "Phys. Rev. Lett.",
    volume = "133",
    number = "22",
    pages = "221003",
    year = "2024"
}

@article{Arteaga:2024vde,
    author = "Arteaga, Mart{\'\i}n and Ghoshal, Anish and Strumia, Alessandro",
    title = "{Gravitational waves and black holes from the phase transition in models of dynamical symmetry breaking}",
    eprint = "2409.04545",
    archivePrefix = "arXiv",
    primaryClass = "hep-ph",
    doi = "10.1088/1475-7516/2025/05/029",
    journal = "JCAP",
    volume = "05",
    pages = "029",
    year = "2025"
}

@article{Kawana:2022olo,
    author = "Kawana, Kiyoharu and Kim, TaeHun and Lu, Philip",
    title = "{PBH formation from overdensities in delayed vacuum transitions}",
    eprint = "2212.14037",
    archivePrefix = "arXiv",
    primaryClass = "astro-ph.CO",
    doi = "10.1103/PhysRevD.108.103531",
    journal = "Phys. Rev. D",
    volume = "108",
    number = "10",
    pages = "103531",
    year = "2023"
}

@article{Gouttenoire:2023naa,
    author = "Gouttenoire, Yann and Volansky, Tomer",
    title = "{Primordial black holes from supercooled phase transitions}",
    eprint = "2305.04942",
    archivePrefix = "arXiv",
    primaryClass = "hep-ph",
    doi = "10.1103/PhysRevD.110.043514",
    journal = "Phys. Rev. D",
    volume = "110",
    number = "4",
    pages = "043514",
    year = "2024"
}

@article{Jung:2021mku,
    author = "Jung, Tae Hyun and Okui, Takemichi",
    title = "{Primordial black holes from bubble collisions during a first-order phase transition}",
    eprint = "2110.04271",
    archivePrefix = "arXiv",
    primaryClass = "hep-ph",
    reportNumber = "KEK-TH-2350",
    doi = "10.1103/PhysRevD.110.115014",
    journal = "Phys. Rev. D",
    volume = "110",
    number = "11",
    pages = "115014",
    year = "2024"
}

@article{Cyburt:2015mya,
    author = "Cyburt, Richard H. and Fields, Brian D. and Olive, Keith A. and Yeh, Tsung-Han",
    title = "{Big Bang Nucleosynthesis: 2015}",
    eprint = "1505.01076",
    archivePrefix = "arXiv",
    primaryClass = "astro-ph.CO",
    reportNumber = "UMN-TH-3432-15, FTPI-MINN-15-19",
    doi = "10.1103/RevModPhys.88.015004",
    journal = "Rev. Mod. Phys.",
    volume = "88",
    pages = "015004",
    year = "2016"
}

@article{Sabti:2019mhn,
    author = "Sabti, Nashwan and Alvey, James and Escudero, Miguel and Fairbairn, Malcolm and Blas, Diego",
    title = "{Refined Bounds on MeV-scale Thermal Dark Sectors from BBN and the CMB}",
    eprint = "1910.01649",
    archivePrefix = "arXiv",
    primaryClass = "hep-ph",
    reportNumber = "KCL-2019-75",
    doi = "10.1088/1475-7516/2020/01/004",
    journal = "JCAP",
    volume = "01",
    pages = "004",
    year = "2020"
}

@article{Pospelov:2010hj,
    author = "Pospelov, Maxim and Pradler, Josef",
    title = "{Big Bang Nucleosynthesis as a Probe of New Physics}",
    eprint = "1011.1054",
    archivePrefix = "arXiv",
    primaryClass = "hep-ph",
    doi = "10.1146/annurev.nucl.012809.104521",
    journal = "Ann. Rev. Nucl. Part. Sci.",
    volume = "60",
    pages = "539--568",
    year = "2010"
}

@article{Graham:2021ggy,
    author = "Graham, Matt and Hearty, Christopher and Williams, Mike",
    title = "{Searches for Dark Photons at Accelerators}",
    eprint = "2104.10280",
    archivePrefix = "arXiv",
    primaryClass = "hep-ph",
    doi = "10.1146/annurev-nucl-110320-051823",
    journal = "Ann. Rev. Nucl. Part. Sci.",
    volume = "71",
    pages = "37--58",
    year = "2021"
}

\end{document}